\documentclass[eqsecnum,flushrt,preprint,10pt]{aastex}

\usepackage{graphicx}
\usepackage{mathrsfs}
\usepackage{amsmath}
\usepackage{bm}

\iftrue

\newcommand{\gapp}{\mathrel{\vcenter{\hbox{\tiny\ooalign{\raise
      3.25pt\hbox{$>$}\crcr $\sim$}}}}}

\newcommand{\be}{\begin{equation}}
\newcommand{\ee}{\end{equation}}
\newcommand{\ba}{\begin{eqnarray}}
\newcommand{\ea}{\end{eqnarray}}

\newcommand{\bra}[1]{\left(#1\right)}
\newcommand{\bras}[1]{\left[#1\right]}
\newcommand{\brac}[1]{\left\{#1\right\}}

\newcommand{\forget}[1]{\iffalse#1\fi}
\newcommand{\forgetmenot}[1]{\iftrue#1\fi}

\newcommand{\sdel}{{\mathrm{D}}}

\newcommand{\udot}{{\cal A}}
\newcommand{\uudot}{\dot{u}}
\newcommand{\n}{n}
\newcommand{\N}{N}
\newcommand{\E}{{\cal E}}
\newcommand{\Hc}{{\cal H}}
\newcommand{\lc}{\varepsilon}
\newcommand{\hatn}{a}
\newcommand{\dotn}{\alpha}
\newcommand{\lb}{\{}
\newcommand{\rb}{\}}

\newcommand{\capl}{L}
\newcommand{\minl}{l}

\newcommand{\El}{\mathscr{E}}
\newcommand{\B}{\mathscr{B}}

\newcommand{\sfS}{_{\mathsf{S}}}
\newcommand{\V}{_{\mathsf{V}}}
\newcommand{\T}{_{\mathsf{T}}}

\newcommand{\lB}{{_{^{^{^{_{\!(\ell_{\!\B\!})\!}}}}}}}
\newcommand{\elg}{{_{^{^{^{_{\!(\ell_{\!g\!})}}}}}}}
\newcommand{\lBg}{{_{^{_{^{^{\!(\ell_{\!\B\!} ,\ell_{\!g\!})\!\!}}}}}}}

\newcommand{\cqg}{Class.\ Quantum Grav.}
\newcommand{\rmp}{Rev.\ Mod.\ Phys.}
\newcommand{\pr}{Phys.\ Rev.}
\newcommand{\prsla}{Proc.\ R.\ Soc.\ London A}

\singlespace

\begin{document}
\baselineskip=0.87\baselineskip
\parskip=0.0\parskip

\title{THE ELECTROMAGNETIC SIGNATURE OF BLACK HOLE RINGDOWN}

\author{C.~A.~Clarkson,\altaffilmark{1,2,5} M.~Marklund,\altaffilmark{3,6}
  G.~Betschart,\altaffilmark{1,3,7} and
  P.~K.~S.~Dunsby\altaffilmark{1,4,8}\\[2mm]\small(\textsc{{\today}})\vspace{-3mm}}

\altaffiltext{1}{Relativity and Cosmology Group,
  Department of Mathematics and Applied Mathematics,
  University of Cape Town, Rondebosch 7701, Cape Town, South Africa\vspace{-2mm}}
\altaffiltext{2}{Institute of Cosmology and Gravitation, University of
  Portsmouth, Portsmouth, PO1 2EG, Britain\vspace{-2mm}}
\altaffiltext{3}{Nonlinear Electrodynamics Group,
  Department of Electromagnetics,
  Chalmers University of Technology, \mbox{SE-412~96} G\"oteborg,
  Sweden\vspace{-2mm}}
\altaffiltext{4}{South African Astronomical Observatory, Observatory
7925, Cape Town, South Africa\vspace{-2mm}}
\altaffiltext{5}{\vspace{-2mm}\texttt{chris.clarkson@port.ac.uk}}
\altaffiltext{6}{\vspace{-2mm}\texttt{marklund@elmagn.chalmers.se}}
\altaffiltext{7}{\vspace{-2mm}\texttt{elfgb@elmagn.chalmers.se}}
\altaffiltext{8}{\vspace{-2mm}\texttt{peter@vishnu.mth.uct.ac.za}}

\forget{
\author{C.~A.~Clarkson}
\email{chris.clarkson@port.ac.uk}
\affil{Relativity and Cosmology Group,
  Department of Mathematics and Applied Mathematics,
  University of Cape Town, Rondebosch 7701, Cape Town, South Africa}
\affil{Institute of Cosmology and Gravitation, University of
Portsmouth, Portsmouth, PO1 2EG, Britain}

\author{M.~Marklund}
\email{marklund@elmagn.chalmers.se}
\affil{Nonlinear Electrodynamics Group,
  Department of Electromagnetics,
  Chalmers University of Technology, SE-412 96 G\"oteborg, Sweden}

\author{G.~Betschart}
\email{geroldb@maths.uct.ac.za}
\affil{Relativity and Cosmology Group,
  Department of Mathematics and Applied Mathematics,
  University of Cape Town, Rondebosch 7701, Cape Town, South Africa}
\affil{Nonlinear Electrodynamics Group,
  Department of Electromagnetics,
  Chalmers University of Technology, SE-412 96 G\"oteborg, Sweden}

\author{P.~K.~S.~Dunsby}
\email{peter@vishnu.mth.uct.ac.za}
\affil{Relativity and Cosmology Group,
  Department of Mathematics and Applied Mathematics,
  University of Cape Town, Rondebosch 7701, Cape Town, South Africa}
}

\begin{abstract}
\baselineskip=0.87\baselineskip

We investigate generation of electromagnetic radiation by
gravitational waves interacting with a strong magnetic field in
the vicinity of a vibrating Schwarzschild black hole. Such an
effect may play an important role in gamma-ray bursts and
supernovae, their afterglows in particular. It may also provide an
electromagnetic counterpart to  gravity waves in many situations
of interest, enabling easier extraction and verification of
gravity wave waveforms from gravity wave detection. We set up the
Einstein-Maxwell equations for the case of odd parity gravity
waves impinging on a static magnetic field as a covariant and
gauge-invariant system of differential equations which can be
integrated as an initial value problem, or analysed in the
frequency domain. We numerically investigate both of these cases.
We find that the black hole ringdown process can produce
substantial amounts of electromagnetic radiation from a dipolar
magnetic field in the vicinity of the photon sphere.
\end{abstract}
\keywords{black hole physics --- gravitational waves --- magnetic fields}


\section{Introduction}

In recent years there has been an enormous effort worldwide to detect
gravitational radiation (see,
e.g.,~\citet{LIGO1,LIGO2,LIGO3,LIGO4}). It is hoped that within the
next few years these detectors will be able to consistently detect and
measure the gravity waves (GW) emitted from such events as black hole (BH)
merger~\citep{buonanno} and exploding and collapsing stars. A pressing problem
for these detectors is the extraction of the actual waveform from the huge
amount of noise invariably generated in the detection process. The race is
currently on to calculate these waveforms in every conceivable situation in
order that gravity wave signatures can eventually be statistically extracted
from the noise continuously generated by these
detectors~\citep{extraction,Flanagan,Nicholson-Vecchio98}, a
formidable task. We discuss here a mechanism describing how many of these
events could be accompanied by an electromagnetic (EM) counterpart with the
same waveform, which could considerably aid in this process.


Many events will be accompanied by an optical counterpart, such as in
supernovae (SN) II and some compact binary mergers~\citep{syl}, but many in
general will not, such as BH-BH merger, and BH ringdown. In any case these will
only tell us to expect detection, and not the precise form of the waveform to
try to extract. What would be highly useful for GW detection would be a
simultaneous optical detection of the event with the EM waveform mirroring that
of the GW. This is the situation we discuss here.

When a plane gravity wave passes through a magnetic field, it vibrates the
magnetic field lines, thus creating EM radiation with the same frequency as
the forcing GW, an effect which has been known for some time (see, e.g.,
~\citet{EM-GW,Gerlach,MBD} and references therein). This would provide
exactly the
mechanism required: virtually all stars have a strong magnetic field
threading through and surrounding them, and this field becomes immensely
strong as the field lines are compressed as the star collapses to a BH or
neutron star; anything up to $10^{14}$G~-- possibly even higher~-- seems
possible in magnetars~\citep{magnetar}.

It has been proposed that this mechanism may indeed have been observed, being
partly responsible for the afterglow observed in some gamma-ray bursts (GRB)
and SN events, an argument strengthened by certain anomalous GRB and SN light
curves (see~\citet{cuesta} and references therein for a detailed discussion).
The basic idea is that these events are thought to form a BH or neutron star
after the initial explosion (envelope ejection), surrounded by a thin plasma
which can support a strong magnetic field able to reach supercritical values
over a relatively long period of time (compared to the period of the emitted
GW). The formation of the compact object will release a substantial fraction
of its mass as GW which could then be converted into EM radiation as it
passes through the plasma. Individual models differ considerably in many
respects; in particular, additional GW may be produced from a stressed
accretion disk powered by the spin of the BH~\citep{putten}.

Studies of generation of EM radiation by GW in astrophysical situations so
far have provided order of magnitude estimates~\citep{cuesta} and some of the
extra complexities involved when a thin plasma is
present~\citep{Macedo-Nelson83,DT,PRL,MBD,Servin-Brodin-Marklund01}.
In particular, a thin plasma can increase the frequency of the
electromagnetic radiation, whose origin
is from a plane gravity wave passing through a uniform, static, magnetic
field, thus strengthening the observational potential of the EM-GW
interaction still further~\citep{BMS,SB}. While these investigations have
given a good indication of the physical processes we may expect, the effect
has not yet been studied in a strong gravitational field, the most promising
place we may expect such an interaction as likely to happen.

Our aim here, then, is to study the induced EM field from the interaction of
GW emitted during BH ringdown, the settling down of a BH after an initial
perturbation, with a strong magnetic field which surrounds the BH. Shortly
after a BH is disturbed by any kind of small perturbation, it radiates its
curvature deformations as GW with certain characteristic frequencies which
are independent of the initial perturbation, and dependent only on its mass
(in the case of a Schwarzschild BH, which we consider here). These complex
quasi-normal frequencies form solutions known as quasi-normal modes which
govern the BH ringdown process~\citep{QNM,Kokkotas-Schmidt99}. As the
ringdown process is thought
to be independent of the initial perturbation, we may expect that studying
this particular situation will help give understanding in more complex
situations, such as the late stages of BH-BH merger~\citep{buonanno}. Indeed,
while it would seem logical that perturbations of Schwarzschild would give
little information about something as non-linear as colliding black holes, it
turns out that perturbation theory gives surprising accuracy in many cases of
interest~\citep{PP}.

The frequency of this generated EM radiation will be very low, generally less
than about $100$kHz, and would be typically absorbed by the interstellar
medium. This is where the photon frequency conversion~\citep{MBD,BMS,SB} could
come into play, overcoming this by increasing the frequency to detectable
levels. An important extension of this work, therefore, will be to include a
plasma into this situation; we leave this to later, and concentrate here on
setting up a suitable formalism for the inclusion of a plasma, while
investigating the pure curvature effects of the BH, which turn out to be
quite large. We find that the amplification of the EM field is much stronger
than is the case of plane GW~\citep{MBD}, where the amplification grows
linearly with interaction distance. Here, we find substantial growth in the
vicinity of the horizon and photon sphere.

\subsection{An overview}

Electromagnetic waves  around a Schwarzschild BH generated by gravitational
waves interacting with a strong static magnetic field are governed by the
Einstein-Maxwell equations which are of the form
\ba
\mbox{EM field around a BH}
=\bra{
\mbox{{GW}}\times\mbox{{Strong static magnetic field}}}.
\label{overview}
\ea
The homogeneous solution to these equations, where the induced currents are
zero, will be governed by the well known Regge-Wheeler (RW) equation for EM
perturbations around a BH~\citep{price72}, while the GW terms are governed by
the Regge-Wheeler equation for gravitational perturbations of a BH~\citep{RW}.
The RW equation describes how the fields of different spin are lensed and
scattered around the hole. A description of the more general situation
including the rhs of Eq.~(\ref{overview}) is rather less trivial than
Eq.~(\ref{overview}) may suggest, for several reasons: for example, the GW
terms on the rhs need some manipulation to convert them to the familiar RW
variable, and the existence of the magnetic field on both sides of the equation
means that gauge problems are paramount, and considerable work must be done to
cast the equations into a manifestly gauge-invariant form. While it may be
expedient to use the Newman-Penrose formalism~\citep{chandra} for this problem
as all variables in the perturbed spacetime travel on null cones, an important
extension of this work will be to include plasma effects, to model a more
realistic astrophysical environment, for which the Newman-Penrose formalism is
not so adept. In addition, the electric and magnetic fields require a timelike
vector field for their definition. For these reasons we will use the covariant
and gauge-invariant perturbation method introduced in~\citet{CB}, which is
ideally suited to this particular problem involving spherical symmetry, and has
the advantage that is is well adapted to fluids for later use when
investigating plasma effects.

An important issue in relativistic perturbation theory is the mapping, or
gauge choice,  one makes between the background and perturbed model; many
perturbation approaches are not invariant with respect to this gauge choice.
Metric based  approaches to perturbation theory suffer from this gauge
freedom, whereby spurious gauge modes exist and must be identified. While
these spurious gauge modes may be eliminated in analytical treatments, when
the equations are integrated numerically these modes have a tendency to grow
very fast without bound, in so-called gauge pathologies~\citep{AM}.
Furthermore, the tractability of the problem often depends upon judicious
gauge choice; hardly an ideal situation~\citep{RSK}.

The covariant `1+1+2' approach we utilise relies on the introduction of a
partial frame which form the differential operators of the spacetime, and allow
all objects to be split into invariantly defined physical or geometric objects.
Covariant perturbation techniques initiated in~\citet{BE} are employed to write
the equations in a fully gauge-invariant form which can then be solved with the
use of appropriate harmonic functions which remove the tensorial nature of the
equations. To aid in the solution we consider it formally as a second-order
perturbation problem, and introduce `interaction variables' for quadratic
quantities. This then allows us to write the equations for the induced EM
radiation as a system of gauge-invariant, covariant,  first order ordinary
differential equations in the relevant variables, while we can easily convert
to covariant wave equations for clarity and integration as an initial value
problem when desired.

The paper is organised as follows. The covariant formalism we use
is reviewed in Sec.~\ref{sec2}. In Sec~\ref{sec3} we derive the
coupled perturbation equations which govern the interaction.
These equations are integrated numerically in Sec.~\ref{sec4}, and
the implications for the emitted radiation discussed. We briefly
conclude in Sec.~\ref{sec5}. An Appendix gives some key formulae
relating to the spherical harmonics we use.

Sections~\ref{sec2} and~\ref{sec3} contain most of the technical
material, the crucial results being Eqs.~(\ref{wave1})~--
(\ref{sources}) and (\ref{alleqnsevenodd}); some readers may wish
to skip to Section~\ref{sec4} where specific astrophysical
situations are discussed.

\section{The 1+1+2 covariant approach}\label{sec2}


Covariant methods in General Relativity (GR) are formulated in a very
different way from coordinate metric based approaches: in the latter,
Einstein's field equations (EFE) are second order partial differential
equations in the components of the metric; in the former, a physical
(partial) frame is chosen and the Ricci and Bianchi identities are
irreducibly split with respect to this frame, resulting in a system of
first-order differential equations. Supplemented by the crucial commutation
relations between the frame vectors, this system of equations becomes
equivalent to  the EFE, but always deals with invariantly defined physical or
geometric quantities.

The 1+1+2 covariant sheet approach relies on the introduction of two frame
vectors: the first being a timelike vector field $u^a$:~$u^au_a=-1$,
representing the congruence on which observers sit; the second is a spacelike
vector field $\n^a$:~$\n^a\n_a=1,~\n^au_a=0$, which can be chosen along a
preferred direction of the spacetime. These two vector fields define a
projection tensor
\be
\N_a^{~b}\equiv h_a^{~b}-\n_a\n^b=g_{a}^{~b}+u_au^b-\n_a\n^b,
\ee
which projects vectors orthogonal to $\n^a$ and $u^a$:
$\n^a\N_{ab}=0=u^a\N_{ab}$, onto 2-surfaces ($\N_a^{~a}=2$) which we refer to
as the `sheet'; $h_{ab}=g_{ab}+u_au_b$ is the tensor which projects orthogonal
to $u^a$, into the observers' rest space. Using $h_{ab}$, any 4-vector may be
split into a (1+3 scalar) part parallel to $u^a$ and a (1+3-vector) part
orthogonal to $u^a$. Any second rank tensor may be covariantly and irreducibly
split into scalar, 3-vector and \emph{projected, symmetric, trace-free (PSTF)}
3-tensor parts, which requires the alternating tensor
$\lc_{abc}=u^d\eta_{dabc}$; these are the key quantities in the 1+3 covariant
approach~\citep{HvE}. Crucially, the covariant derivative of $u^a$ may be split
in the standard manner, the irreducible parts (the acceleration $\uudot^a$, the
expansion, $\theta$, the shear, $\sigma_{ab}$, and the rotation, $\omega^a$)
forming some of the key variables of the 1+3 approach~-- we refer
to~\citet{HvE} for further details.

Now, using $N_{ab}$, any 3-vector $\psi^a$ can now be irreducibly split into a
scalar, $\Psi$, which is the part of the vector parallel to $\n^a$, and a
2-vector, $\Psi^a$, lying in the sheet orthogonal to $\n^a$:
\begin{mathletters}
\be
\psi^a = \Psi\n^a + \Psi^{a},
\ee
where
\be
\Psi\equiv \psi_a\n^a,~~~\mbox{and}~~~\Psi^{a}\equiv
\N^{ab}\psi_b\equiv \psi^{\bar a},
\ee
\label{vector-decomp}
\end{mathletters}
where we use a bar over an index to denote projection with $\N_{ab}$.
Similarly, any PSTF tensor, $\psi_{ab}$, can now be split into scalar, vector
and 2-tensor (which are PSTF with respect to $\n^a$, and is therefore
transverse-traceless) parts:
\begin{mathletters}
\be
\psi_{ab} = \psi_{\langle ab\rangle } = \Psi\bra{\n_a\n_b - \tfrac{1}{2}\N_{ab}} +
2\Psi_{(a}\n_{b)} + \Psi_{{ab}},
\ee
where
\ba
\Psi&\equiv &\n^a\n^b\psi_{ab}=-\N^{ab}\psi_{ab}, \\
\Psi_a&\equiv &\N_a^{~b}\n^c\psi_{bc}=\Psi_{\bar a}, \\
\Psi_{ ab}&\equiv &\psi_{\lb ab\rb}\equiv
\bra{\N_{(a}^{~~c}\N_{b)}^{~~d}-\tfrac{1}{2}\N_{ab}\N^{cd}}\psi_{cd}.\label{tensor-decomp}
\ea
\label{PSTF-TT}
\end{mathletters}
We use curly brackets to denote the transverse-traceless (TT) part of a
tensor. We also define the alternating Levi-Civita 2-tensor (area element)
\be
\lc_{ab}\equiv\lc_{abc}\n^c = u^d\eta_{dabc}n^c,
\ee
so that $\lc_{ab}\n^b=0=\lc_{(ab)}$.

With these definitions, then, we may split any object into \emph{scalars,
2-vectors in the sheet, and transverse-traceless 2-tensors, also defined in
the sheet.} These three types of objects are the only objects which need to
be solved for, after a complete splitting. Hereafter, we will assume such a
split has been made, and `vector' will generally refer to a vector projected
orthogonal to $u^a$ and $\n^a$, and `tensor' will generally mean
transverse-traceless tensor, defined by Eq.~(\ref{PSTF-TT}).

We split the familiar 1+3 variables in this manner; in particular, the electric
and magnetic fields are irreducibly split
\begin{mathletters}
\ba
E^a &=& \El\n^a + \El^a,\\
B^a &=& \B \n^a + \B^a,
\ea
\end{mathletters}
while the kinematical and gravitational variables become, using
Eqs.~(\ref{vector-decomp}) and~(\ref{PSTF-TT})
\begin{mathletters}
\ba
\uudot^a &=& \udot \n^a + \udot^a,\\
\omega^a &=& \Omega \n^a + \Omega^a,\\
\sigma_{ab} &=& \Sigma\bra{\n_a\n_b - \tfrac{1}{2}\N_{ab}} +
2\Sigma_{(a}\n_{b)} + \Sigma_{ab},\\
E_{ab} &=& {\cal E}\bra{\n_a\n_b - \tfrac{1}{2}\N_{ab}} + 2{\cal
E}_{(a}\n_{b)} + {\cal E}_{ab},\\
H_{ab} &=& {\cal H}\bra{\n_a\n_b - \tfrac{1}{2}\N_{ab}} + 2{\cal
H}_{(a}\n_{b)} + {\cal H}_{ab}.
\ea
\end{mathletters}
For example, $\E=n^an^bE_{ab}$ is the tidal force along $\n^a$,
$\E_a=\N_a^{~b}\n^cE_{bc}$ is a 'drift' vector, while $\E_{ab}=E_{\{ab\}}$ is
the TT part of the electric Weyl curvature in the sheet orthogonal to $\n^a$.

There are two new derivatives of interest, which $\n^a$ defines, for any object
$\psi_{\cdots}^{~~\cdots}$:
\begin{mathletters}
\ba
\!\!\!\!\!\!\!\!\!\hat \psi_{a\cdots b}^{~~~~~c\cdots d} &\equiv &
\n^e \sdel_e\psi_{a\cdots b}^{~~~~~c\cdots d},\label{hatdef}\\
\!\!\!\!\!\!\!\!\!\delta_e \psi_{a\cdots b}^{~~~~~c\cdots d}&\equiv & \N_e^{~j}\N_a^{~f}\cdots
\N_b^{~g}\N_h^{~c}\cdots\N_i^{~d}\sdel_j \psi_{f\cdots g}^{~~~~~h\cdots i},\label{deltadef}
\ea
\end{mathletters}
where $\mathrm{D}_a$ is the spatial derivative defined by $h_a^{~b}$ (see,
e.g., ~\citet{HvE})
.  The
hat-derivative is the derivative along the vector field $\n^a$ in the surfaces
orthogonal to $u^a$. It is important to note, however, that these derivatives
do not commute; commutation relations for scalars are given in~\citet{CB}. This
is a vital aspect of the formalism.

With these definitions we may now decompose the spatial projection of the
covariant derivative of $\n^a$ orthogonal to $u^a$:
\be
\sdel_a\n_b=\n_a\hatn_b+\tfrac{1}{2}\phi \N_{ab}+\xi\lc_{ab}+\zeta_{ab},
\ee
where
\begin{mathletters}
\ba
\hatn_a &\equiv &\n^c\sdel_c\n_a=\hat \n_a,\\
\phi &\equiv &\delta_a \n^a,\\
\xi &\equiv &\tfrac{1}{2}\lc^{ab}\delta_a\n_b,\\
\zeta_{ab} &\equiv &\delta_{\lb a}\n_{b\rb}.
\ea
\end{mathletters}
We may interpret these as follows: travelling along $\n^a$, $\phi$ represents
the sheet expansion, $\zeta_{ab}$ is the shear of $\n^a$ (distortion of the
sheet), and $\hatn^a$ its acceleration, while $\xi$ represents a `twisting' of
the sheet~-- the rotation of $\n^a$. The other derivative of $\n^a$ is its
change along $u^a$,
\be
\dot n_{a}=\udot u_a+\dotn_a~~~\mbox{where}~~~\dotn_a\equiv\dot n_{\bar a}
~~~\mbox{and}~~~\udot=\n^a\uudot_a.
\ee
The new variables $\hatn_a$, $\phi$, $\xi$, $\zeta_{ab}$ and $\dotn_a$ are
fundamental objects in the spacetime, and their dynamics gives us information
about the spacetime geometry. They are treated on the same footing as the
kinematical variables of $u^a$ in the 1+3 approach (which also appear here).

The 1+1+2 split of the Ricci identities for $u^a$ and $\n^a$, and the Bianchi
identities, provide a complete set of first-order differential equations for
these variables, and were discussed in~\citet{CB} for the case of a
gravitationally perturbed BH. We shall not require a further generalisation of
these equations here.

Obviously, we shall require Maxwell's equations (ME), which  may be
irreducibly split using these definitions:
\begin{mathletters}
\ba
 \hat \El +\delta_a \El ^a &=&   -\phi \El  + \El _a\hatn^a +2\Omega \B
+2\Omega^a \B _a + \mu_0 \rho_{\mathrm{e}},\\
 \hat \B +\delta_a \B ^a &=&   -\phi \B  + \B _a\hatn^a -2\Omega \El
-2\Omega^a \El _a,\label{MEgen1}\\
 \dot \El -\lc_{ab}\delta^a \B ^b &=&   2\xi
\B +\El ^a\dotn_a-\bra{\tfrac{2}{3}\theta-\Sigma} \El
+\Sigma^a \El _a +\lc_{ab}\bra{\udot^a \B ^b +\Omega^a \El ^b}-
\mu_0{\cal J},\\
 \dot \B +\lc_{ab}\delta^a \El ^b &=&-2\xi
\El +\B ^a\dotn_a-\bra{\tfrac{2}{3}\theta-\Sigma} \B
+\Sigma^a \B _a -\lc_{ab}\bra{\udot^a \El ^b -\Omega^a \B ^b},\\
 \dot \El _{\bar a}+\lc_{ab}\bra{\hat \B ^b-\delta^b \B } &=&    \xi \B _a
-\bra{\tfrac{1}{2}\phi+\udot}\lc_{ab}\B ^b
-\bra{\tfrac{2}{3}\theta+\tfrac{1}{2}\Sigma}\El _a-\Omega\lc_{ab}\El
^b\nonumber\\&&
 + \El \bra{-\dotn_{a}+\Sigma_a+\lc_{ab}\Omega^b} +\B
\lc_{ab}\bra{\udot^b-\hatn^b}
+\Sigma_{ab}\El ^b-\lc_{ab}\zeta^{bc}\B _c-\mu_0{\cal J}_a,\label{MEgen5}\\
 \dot \B _{\bar a}-\lc_{ab}\bra{\hat \El ^b-\delta^b \El } &=& -\xi \El _a
+\bra{\tfrac{1}{2}\phi+\udot}\lc_{ab}\El ^b
-\bra{\tfrac{2}{3}\theta+\tfrac{1}{2}\Sigma}\B _a-\Omega\lc_{ab}\B
^b\nonumber\\&&
 + \B \bra{-\dotn_{a}+\Sigma_a+\lc_{ab}\Omega^b} -\El
\lc_{ab}\bra{\udot^b-\hatn^b}
+\Sigma_{ab}\B ^b+\lc_{ab}\zeta^{bc}\El _c.\label{MEgen}
\ea
\end{mathletters}
Here, MKS units are used ($\mu_0$), $\rho_e$ is the charge density, and the
current density $j_a$ has been split into its $1+1+2$ parts, ${\cal J}$ and
${\cal J}_a$. The first two equations arise from the constraint ME,
while the rest
are the evolution ME. In flat space in the absence of currents and
charges the rhs of these equations vanish (for a static `natural' choice of
frame). Thus, gravity modifies ME in the form of generalised currents. Note
how the rotation terms $\xi,~\Omega$ and $\Omega^a$ flip the parities of the
EM fields ($\lc_{ab}$ is a parity operator~-- see the Appendix).

\section{Electromagnetic radiation around a vibrating black
hole}\label{sec3}

There are two ways to proceed in solving this problem, depending on how one
views Maxwell's and Einstein's equations. If we view Maxwell's equations as
being essentially separate from the field equations, deciding on the fly
whether to include the gravitational effects of the EM field, then this
particular situation may be considered as having the EM field as a test field
on a vibrating BH background. An alternative viewpoint is to consider Maxwell's
and Einstein's equations  as a \emph{coupled} system of equations (with
$\mu_B\sim\tfrac{1}{2}B^2$, etc.), with decoupling occurring only when one can
legitimately set terms ${\cal O}(B^2)$ to zero, clearly more intuitive in a
perturbation approach. They are mathematically equivalent in vacuo only due to
the linearity of ME, a feature not present in plasmas in general.\footnote{We
also consider the second interpretation more appropriate in a covariant
approach because the frame derivatives in ME automatically couple the
\emph{curvature} of the spacetime to the EM field through the commutation
relations [for example, $\sdel_{[a}\sdel_{b]}V_c$ brings in both Ricci and Weyl
curvature in general, and we use this commutator after defining gauge-invariant
variables~-- see Eq.~(\ref{hatbeta_a}) below]. On the other hand, the coupling
to gravity which occurs \emph{explicitly} in ME in any curved spacetime is from
purely \emph{kinematical} quantities (frame motion), which then couple to
gravity through the field equations;  this is why gravity can be explicitly
`decoupled' from the vacuum ME. The commutation relations thus confuse this
issue, because through these relations curvature directly induces the EM
field.} By using the second interpretation things will automatically be easier
when complicated non-linear plasma effects are included at a later date. We
will therefore treat this as a perturbation problem at \emph{second-order} in a
two parameter `expansion' in two `smallness' parameters $\epsilon_\B$
representing the magnitude of the static magnetic field, and $\epsilon_g$
representing the amplitude of the GW (these are labels for the two types of
first-order perturbations as much as anything else; we need both because we
keep terms ${\cal O}(\epsilon_\B\epsilon_g)$, but neglect terms ${\cal
O}(\epsilon_g^2)$ and ${\cal O}(\epsilon_\B^2)$)~--
see~\citet{BS,Bruni-Sonego99}. Consequently, this expansion allows us to set up
the equations as a system of \emph{linear} first-order differential equations
at second-order in the perturbation, which at the same time serves to
illustrate a new technique for covariant and gauge-invariant non-linear
perturbation theory. Once a gauge-invariant formalism has been set up to study
this interaction it should then be relatively easy to include complicated
plasma effects and so on.

We will divide up the perturbation `background' spacetimes and denote them as
follows:
\begin{itemize}
\item[--] ${\cal B}=$~Exact Schwarzschild, ${\cal O}(\epsilon^0)$;
\item[--] ${\cal F}_1=$~Exact Schwarzschild perturbed by a pure static magnetic field,
        neglecting the energy density of the field in comparison to the curvature of the BH: ~${\cal
        O}(\epsilon_\B)$;
\item[--] ${\cal F}_2=$~Schwarzschild with gravitational perturbations~${\cal
O}(\epsilon_g)$, as given in~\citet{CB,RW};
\item[--] ${\cal S}={\cal F}_1+{\cal F}_2$ allowing for interaction terms in Maxwell's equations:
    the induced EM fields will be ${\cal O}(\epsilon_\B\epsilon_g)$; this is
    the situation of interest.
\end{itemize}
We will generally refer to terms of order ${\cal O}(\epsilon_\B)$ and ${\cal
O}(\epsilon_g)$ appearing in ${\cal F}$ as `first-order', while those
variables of order ${\cal O}(\epsilon_\B\epsilon_g)$ in ${\cal S}$ as
`second-order' variables. If one prefers to view ME as a test field (where
the only `perturbation' is in the EM field), then these backgrounds may
instead be thought of as useful labels for each type of field present.

\subsection{The background fields}

We now review each of the backgrounds.

\vskip\baselineskip
\noindent \textsc{${\cal B}$: the exact schwarzschild solution.}
For a family of static observers, in the background we have only the
zeroth-order scalars: $\E$, the \emph{radial tidal force}; $\udot$, the
acceleration a static observer must apply radially outwards (to prevent
infall); and $\phi$, the spatial expansion of the radial vector $\n^a$. These
are determined by the radial propagation equations\footnote{We refer to
equations in involving the radial hat-derivative as \emph{propagation
equations}, and those involving the temporal dot-derivative as \emph{evolution
equations}; equations involving neither of these may be thought of as
constraints, though this depends on how one chooses to integrate the
equations~-- see below.}:
\begin{mathletters}
\ba
\hat\phi &=&-\tfrac{1}{2}\phi^2-{\cal E},\label{phihatrad}\\
\hat{\cal E}&=&-\tfrac{3}{2}\phi{\cal E}\label{ehatrad};
\ea
together with
\be
{\cal E}+\udot\phi=0.\label{udotbackground}
\ee
\end{mathletters}
Defining the affine parameter by $\hat{\phantom{x}}=d/d\rho$, and another
radial parameter $r$ by
\be
\hat r=\tfrac{1}{2}\phi r,
\ee
the parametric solution to these equations, giving a complete description of
the BH, are given by
\begin{mathletters}
\ba
{\cal E}&=&-\frac{2m}{r^3},\\
\phi &=&\frac2r\sqrt{1-\frac{2m}{r}},\\
\udot&=&\frac{m}{r^2}\bra{1-\frac{2m}{r}}^{-1/2};
\label{backsols(x)}
\ea
\end{mathletters}
where
\be
\rho = 2m\cosh^{-1}\bra{\sqrt{\frac{r}{2m}}}+{r}\sqrt{1-\frac{2m}{r}},\label{rho(x)}
\ee
relates the affine parameter $\rho$ associated with the radial vector $\n^a$
with the usual Schwarzschild coordinate $r$.\newline \emph{In ${\cal F}$ and
${\cal S}$ we keep all powers of these variables.}

\vskip\baselineskip
\noindent \textsc{${\cal F}_1$: the static magnetic field.}
The following equations govern the static ($\dot \B =\dot \B_{a}= 0$)
magnetic field:
\begin{mathletters}
\ba
\hat \B&=&  -\delta_a \B^a -\phi \B ,\\
\hat \B_a&=& \delta_a \B -\bra{\tfrac{1}{2}\phi+\udot}\B_a,\\
0 &=&\lc_{ab}\delta^a \B^b   .\label{B1eqs}
\ea
\end{mathletters}
The last equation tells us that the field is purely of even parity. In general
the solution to these equations when harmonically decomposed can only be
written as a complicated combination of hypergeometric functions (which is
partly why the perturbation method we utilise below is effective).\newline
\emph{In ${\cal S}$ we neglect all products of the magnetic field with itself.}

The solution for a dipole field is of particular importance: when split into
spherical harmonics (see~\citet{CB} and the Appendix), the $\ell=1$ equations
have two solutions; one which is uniform at infinity, characterised by
$\hat\B=0$ and one which falls off like $1/r^3$ at infinity, which is the true
dipole. The solution for the latter part is, in terms of $r$:
\begin{mathletters}
\ba
\B\sfS &=&
-\frac{3\B_\infty}{8m^3}\bras{\ln\bra{1 - \frac{2m}{r}} +
  \frac{2m}{r}\bra{1 + \frac{m}{r}}},\\
\B\V &=& -\frac{3\B_\infty}{8m^3}\bra{1 - \frac{2m}{r}}^{-1/2}
    \bras{\bra{1 - \frac{2m}{r}}\ln\bra{1 - \frac{2m}{r}} +
      \frac{2m}{r}\bra{1-\frac{m}{r}}},
\ea
\label{Bdipolesol}
\end{mathletters}
where $\B_\infty$ is the magnitude of $\B\sfS r^3$ as $r\rightarrow\infty$.

\vskip\baselineskip
\noindent \textsc{ ${\cal F}_2$: the gravity wave perturbation.}
As shown in~\citet{CB}, these perturbations are governed completely, in the
$\udot^a=\delta^a\phi=\delta^a\udot=0$ frame, by the tensorial form of the
Regge-Wheeler equation \citep{RW,CB}
\ba
-\ddot W_{ab}+\hat{\hat W}_{ab}+\udot{\hat W}_{ab} -\phi^2 W_{ab}+\delta^2
W_{ab} =0,\label{RW}
\ea
where the \emph{Regge-Wheeler tensor} $W_{ab}$ is a gauge- and
frame-invariant TT tensor, defined as \citep{CB}
\be
W_{ab}=\tfrac{1}{2}\phi r^2\zeta_{ab}-\tfrac{1}{3}r^2\E^{-1}\delta_{\lb
a}X_{b\rb},\label{Wab}
\ee
and $X_a=\delta_a\E$ is the gauge-invariant variable~\citep{SW} describing the
angular fluctuation in the radial tidal force. This tensor contains in compact
form the curved space generalisation of the two flat space GW polarizations
$h_+$ and $h_\times$\footnote{In Minkowski space, in the transverse-traceless
gauge, gravitational waves are described by the TT tensor,
\[
h^{TT}_{\mu\nu}=\bra{\begin{array}{cc}
  h_+ & h_\times \\
  h_\times & -h_+
\end{array}}
\]
(see Box 35.1 in~\citet{MTW}). In cartesian coordinates for a plane wave
travelling along the $z$-axis,
$\zeta_{\mu\nu}=\tfrac{1}{2}\partial_zh^{TT}_{\mu\nu}$. Thus, far from a
localised source of radiation, the scale-invariant part of the GW is
$W_{ab}\simeq r\zeta_{ab}$ which is the power of the GW; hence $\langle
W_{ab}W^{ab}\rangle $ (where the angle brackets denote the average over several
wavelengths) is proportional to the energy carried from the source by the
GW~\citep{MTW}.}.

Every other object in ${\cal F}_2$ is determined by linear combinations of
$\{W
\T,\hat W\T,\bar W\T\,\hat{\bar{W}}\T\}$, once appropriate harmonics are
used (see the Appendix and~\citet{CB}). While Eq.~(\ref{RW}) governs GW of both
parities, for simplicity we shall only consider the case here where the GW are
of odd parity. For purely odd perturbations the gravitational field is governed
by $W_{ab}$, and the other GW variables that we shall require are related to
this by the \emph{covariant} gauge-invariant equations~\citep{CB}
\begin{mathletters}
\ba
\zeta_{ab}&=&\frac{2}{\phi r^2}W_{ab}\label{zeta-W},\\
\dot\Sigma_{ab}&=&r^{-2}W_{ab}+\frac{2}{\phi r^2}\hat W_{ab},\label{dot-sig}\\
\dot\alpha_a=-\dot\Sigma_a=\E_a&=&\frac{2}{\phi r^2}\delta^bW_{ab},\label{dot-sig2}\\
\lc_{ab}\dot\Hc^b&=&-\frac{2}{\phi r^2}\delta^b\hat W_{ab},\\
\dot\Hc&=&-\frac{2}{\phi r^2}\lc_{ab}\delta^a\delta^cW^b_{~c}.
\ea
\end{mathletters}
Although they may be given in a similar fashion, we shall not require
$\E_{ab},\Hc_{ab}$. (All other 1+1+2 variables are zero.) While we would not
normally require parts of the Weyl tensor to solve ME, we will need them here
as they arise when generating propagation equations for the gauge-invariant
part of the magnetic field in ${\cal S}$, through the commutation relations. We
shall find that turning ME into a gauge-invariant system at second order
explicitly introduces Weyl curvature into the problem.\newline \emph{In ${\cal
S}$ we neglect all products of these quantities.}

\vskip\baselineskip
Because the background ${\cal B}$ is spherically symmetric, the solutions of
both parts of ${\cal F}$ may be expanded in spherical harmonics. This implies
that we can write
\begin{mathletters}
\ba
\B&=&\sum_{\ell_\B=1}^\infty\B^\lB,\\
\B_a&=&\sum_{\ell_\B=1}^\infty\B_a^\lB,\\
W_{ab}&=&\sum_{\ell_g=2}^\infty W_{ab}^\elg,
\ea
\end{mathletters}
where the $g$ and $\B$ subscripts serve to remind which harmonic indices we
are summing over in each case, a distinction required in the next Section.
Then the harmonic components of the magnetic field and RW tensor obey the
constraint equations, where $L=\ell\bra{\ell+1}$,
\begin{mathletters}
\ba
\delta^2\B^\lB &=& -\capl_\B r^{-2}\B^\lB,\\
\delta^2\B_a^\lB &=& \bra{1-\capl_\B} r^{-2}\B_a^\lB, \\
\lc_{ab}\delta^a\B^b_\lB &=& 0
\ea
\end{mathletters}
and
\begin{mathletters}
\ba
\delta^2 W_{ab}^\elg &=& \bra{\phi^2-3\E-\capl_g r^{-2}}W_{ab}^\elg, \\
\delta^a\delta^bW_{ab}^\elg &=& 0.\label{SHGWB}
\ea
\end{mathletters}
Assuming separable solutions implies the usual spherical harmonics for the
angular parts~-- see the Appendix. We can write the variables in this way
because only one parity is present for each field.

\subsection{The interaction terms in Maxwell's equations}

Here we will introduce a set of auxiliary variables, all of order ${\cal
O}(\epsilon_\B\epsilon_g)$, which allow us to convert ME into a linear (in
differential order) system of \emph{gauge-invariant} ordinary differential
equations (gauge-invariant because they vanish at all perturbative orders
lower than this~\citep{BS,Bruni-Sonego99}). We refer to these as the
interaction variables

A quick glance  at the rhs of ME, Eqs.~(\ref{MEgen1})~-- (\ref{MEgen}),
reveals that we're dealing with products of tensorial spherical harmonics,
which are not particularly pleasant. Instead of explicitly using tensor
spherical harmonics in the GW$\times$B products in ME, we shall absorb them
into the following interaction variables, which makes the resulting equations
considerably neater. There is no extra work involved here, although it may
not appear that way; we would otherwise still require the key
equations~(\ref{chiprop}) and~(\ref{GWBcons}) given below. The latter in
particular are crucial relations among all the coupled tensor/vector/scalar
spherical harmonics which appear (these are the products given by
Eqs~(\ref{chidef}),~(\ref{auxvars}), although we have absorbed the magnetic
field strength and the GW amplitude). There is another reason for defining
the variables in the manner we do: while variables such as $\alpha_a$ appear
in ME, our solution in ${\cal F}_2$ only gives us $\dot\alpha_a$;  we
circumvent this problem by absorbing the time derivatives into our new
variables below.

With these considerations in mind, we define the four \emph{interaction
variables}
\be
\mathbf{\chi}_a^\lBg=\bra{\chi_{1a}^\lBg,\chi_{2a}^\lBg,\chi_{3a}^\lBg,\chi_{4a}^\lBg}
\ee
as follows:
\begin{mathletters}
\ba
\dot\chi_{1a}^\lBg&=&\bra{\phi r^2}^{-1} W_{ab}^\elg\B^b_\lB,\\
\dot\chi_{2a}^\lBg&=&\bra{\phi r^2}^{-1} \hat W_{ab}^\elg\B^b_\lB,\\
\dot\chi_{3a}^\lBg&=&\bra{\phi r^2}^{-1} W_{ab}^\elg\delta^b\B^\lB,\\
\dot\chi_{4a}^\lBg&=& \bra{\phi r^2}^{-1}\hat
W_{ab}^\elg\delta^b\B^\lB,
\ea
\label{chidef}
\end{mathletters}
for each $\ell_\B\leftrightarrow\ell_g$ interaction. We use a bold font as a
matrix shorthand for the `4-vector' these variables form. These variables
obey the propagation equations
\be
\hat{\bm{\chi}}^{\lBg}_a=\bm\Gamma^{\lBg}\bm{\chi}^{\lBg}_a,
\label{chiprop}
\ee
with the \emph{interaction matrices} given by, for each $\ell_\B$ and
$\ell_g$,
\be
\bm\Gamma^\lBg=
\forget{\bra{
  \begin{array}{cc}
    \bm\Lambda^\elg-\bra{\tfrac{1}{2}\phi+\udot}\bm I & \bm I \\
    \capl_\B r^{-2}\bm I & \bm\Lambda^\elg - \tfrac{3}{2}\phi\bm I
  \end{array}}=}
  \bra{\begin{array}{cccc}
  -\bra{\phi + \udot} & 1 & 1 & 0 \\
  \Delta^\elg & -\bra{\phi+2\udot} & 0 & 1 \\
  \capl_\B r^{-2} & 0 & -2\phi & 1 \\
  0 & \capl_\B r^{-2} & \Delta^\elg & -\bra{2\phi+\udot}
\end{array}},
\ee
where
\be
\Delta^\elg\equiv - \omega^2 + 3\E + \capl_g r^{-2}.
\ee
We have introduced the time harmonics of~\citet{CB} into these equations for
notational simplicity; factors of $i\omega$ just represent time derivatives,
$d/d\tau$~-- we will discuss the significance of these later. For now note that
\be
\hat\omega=-\udot\omega~~~\Rightarrow~~~\omega= \sigma\bra{1-\frac{2m}{r}}^{-1/2}=\frac{2\sigma}{\phi r},\label{omdef}
\ee
arising from the commutation relation between the dot- and hat-derivatives.
Here, $\sigma$ is a constant, which we will discuss below.

In order to simplify our presentation, we will define a set of
auxiliary interaction variables as follows (some of
which may be a little surprising, but they are all required). First, define
 \be
\bm V_a^\elg=\bra{\begin{array}{c}
  V_{1a}^\elg \\
  V_{2a}^\elg
\end{array}}=
\bra{\begin{array}{c}
  \delta^bW_{ab}^\elg\\
  \delta^b\hat W_{ab}^\elg
\end{array}}
\ee
where we use a bold font to denote the `2-vector' matrix. Similarly we define
 \be
\bm \lambda_a^\lB=\bra{\begin{array}{c}
  \lambda_{1a}^\lB \\
  \lambda_{2a}^\lB
\end{array}}=
\bra{\begin{array}{c}
  \B_a^\lB\\
  \delta_a \B^\lB
\end{array}}.
\ee
For simplicity of presentation, we introduce the shorthand notation `$\circ$'
which takes two `2-vectors'  to form a `4-vector'  as
\ba
\bm V\circ\bm\lambda=\bra{V_1\lambda_1,V_2\lambda_1,V_1\lambda_2,V_2\lambda_2}.
\ea
We use these to define the following `4-vector' variables as follows
\begin{mathletters}
\ba
\dot{\bm K}^\lBg&=&\bra{\phi r}^{-1}\lc_{ab}\bm
V_{\elg}^a\circ\bm\lambda^b_\lB , \\
\dot{\bm\Psi}_{a}^\lBg&=& ~~ \phi^{-1}~~~~~\bm V_{a}^\elg\circ
\delta^c\bm\lambda_c^\lB,\\
\dot{\bm M}_{a}^\lBg&=&~~\phi^{-1}~~~~~\bm V_{\elg}^b\circ\delta_{\lb
a}\bm\lambda_{b\rb}^\lB,\\
\dot{\bm J}_{a}^\lBg&=&\phi^{-1}\lc_{cd}\delta^c\bm
V_{\elg}^d\circ\lc_{ab}\bm\lambda^{b}_\lB,
\label{auxvars}
\ea
\end{mathletters}
where, for example,
$\bm K=\bra{K_1,K_2,K_3,K_4}$, gives the shorthand for
four of these sixteen new variables. These variables are all ${\cal
O}(\epsilon_\B\epsilon_g)$. They are all constructed to obey the same
propagation equation as $\bm\chi$, viz:
\begin{mathletters}
\ba
\hat{\bm K}^\lBg &=& \bm\Gamma^\lBg\bm K^\lBg,\\
\hat{\bm\Psi}^\lBg_a &=&\bm\Gamma^\lBg\bm\Psi^\lBg_a,\\
\hat{\bm M}^\lBg_a&=&\bm\Gamma^\lBg\bm M^\lBg_a,\\
\hat{\bm J}^\lBg_a&=&\bm\Gamma^\lBg\bm J^\lBg_a.
\ea
\label{GWBprop}
\end{mathletters}
We have defined all these variables as the time integral of combinations of
the RW tensor and the static magnetic field. This is because for some of the
GW variables appearing in ME it's their time derivatives which are related to
the RW tensor [see, e.g., Eqs.~(\ref{dot-sig}) and~(\ref{dot-sig2})].
Defining auxiliary interaction variables this way which satisfy the
propagation equations~(\ref{GWBprop}) removes this problem, and absorbs it
into the initial (or boundary) conditions.

By taking various $\delta$-derivatives of these variables and using the
appropriate commutation relations (see~\citet{CB}), together with
Eq.~(\ref{SHGWB}), we can show that they all obey the following constraints,
which are crucial identities for consistency of the resulting equations later,
and allow us to relate all the interaction terms to $\bm\chi$ when we split ME
into spherical harmonics.
\begin{mathletters}
\ba
0&=&\bm J_a^\lBg+2\bm M_a^\lBg+2r\lc_{ab}\delta^b\bm
    K^\lBg 
+\bra{\capl_g-2}\bm\chi_a^\lBg-\bm\Psi_a^\lBg, \\
0&=& r\delta^2\bm K^\lBg+\lc_{ab}\delta^a\bm\Psi^b_\lBg 
-\bra{\capl_g-2}\lc_{ab}\delta^a\bm\chi^b_\lBg,\\
0&=& \bras{r\delta^2+\bra{\capl_g-\capl_\B}r^{-1}}\bm K^\lBg
+\lc_{ab}\delta^a\bm\Psi^b_\lBg-2\lc_{ab}\delta^a\bm M^b_\lBg, \\
0&=& \capl_\B\bras{\delta^2+\bra{\capl_\B-\capl_g}r^{-2}}\bm K^\lBg
-r\bras{\delta^2+\bra{\capl_\B+\capl_g}r^{-2}}\lc_{ab}\delta^b\bm\Psi^a_\lBg,
    \\
 0&=&
\bra{r^2\delta^2+\capl_\B}\delta^a\bm\Psi_a^\lBg
    -\bra{\capl_g-2}\capl_\B\delta^a\bm\chi_a^\lBg, \\
0&=&
\bra{r^2\delta^2+\capl_g}\delta^a\bm\Psi_a^\lBg -2\capl_g\delta^a\bm
M_a^\lBg.
\ea
\label{GWBcons}
\end{mathletters}
These twenty-four constraint equations propagate consistently.

For each $\ell_\B \leftrightarrow \ell_g$ interaction, the system of equations
describing \emph{the gravitational wave~-- magnetic field interaction} are
given above. Not all these variables appear explicitly in ME, but they couple
to them through the system of propagation equations~(\ref{GWBprop}) and
constraints~(\ref{GWBcons}). We now discuss how these enter ME. Consider, for
example, the term $\zeta_{ab}\B^b$ which appears in the evolution equation for
$\El_a$, Eq.~(\ref{MEgen5}). We can relate this to the interaction variables
above as follows: using Eq.~(\ref{zeta-W}) we have
\be
\zeta_{ab}\B^b=\frac{2}{\phi r^2}\bra{\sum_{\ell_g}W_{ab}^\elg }\bra{\sum_{\ell_\B}B^b_\lB}
=2\sum_{\ell_\B,\ell_g}\dot\chi_{1a}^\lBg.
\ee
Similarly for the other products. We therefore use the following
abbreviations:
\begin{eqnarray}
\bm K=\sum_{\ell_\B,\ell_g}\bm K^\lBg,~~~
\bm\chi_a=\sum_{\ell_\B,\ell_g}\bm\chi_a^\lBg,~~~
\bm M_a=\sum_{\ell_\B,\ell_g}\bm M_a^\lBg,~~~
\bm J_a=\sum_{\ell_\B,\ell_g}\bm J_a^\lBg,
\end{eqnarray}
while for $\bm\Psi_a$ we define:
\be
\bm\Psi_a =\sum_{\ell_\B,\ell_g}
\bra{
   \Psi_{1a}^\lBg,
   \Psi_{2a}^\lBg,
   L_\B^{-1}\Psi_{3a}^\lBg,
   L_\B^{-1}\Psi_{4a}^\lBg }.
\ee
(The definition for $\bm\Psi$ is slightly different because it is defined
having a $\delta^2\B^\lB\sim\capl_\B\B^\lB$ term in it.) These definitions
prevent summations appearing explicitly later.

\subsection{The gauge-invariant form of Maxwell's equations}

Neglecting terms ${\cal O}\bra{\epsilon_\B\times\mbox{even parity gravity
waves}}$ and ${\cal O}(\epsilon_\B^2)$ (strictly speaking we are only
neglecting terms ${\cal O}(\epsilon_\B^3)$, see the end of this
section), and choosing the frame in ${\cal
F}_2$ such that $\udot^a=\delta^a\phi=\delta^a\udot=0$ we find that ME
become,
\begin{mathletters}
\ba
\hat{\El} + \delta_a \El^a+\phi \El &=& 0,\label{Eshat}\\
\hat{\B} + \delta_a \B^a +\phi \B &=& 0\label{hatB2} ,\\
\dot{\El} - \lc_{ab}\delta^a \B^b &=&  0,\label{Esdot}\\
\dot{\B} + \lc_{ab}\delta^a \El^b &=&0,\label{BSdot}
\ea
and
\ba
&&\dot{\El}_{\bar{a}} + \lc_{ab}\bra{\hat{\B}^b - \delta^b \B}
+
\bra{\tfrac{1}{2}\phi + \udot}\lc_{ab}\B^b =
-2\lc_{ab}\dot{\chi}_1^b,\label{Bahat}\\
&&\dot{\B}_{\bar{a}}
 -\lc_{ab}\bra{\hat{\El}^b - \delta^b \El}
- \bra{\tfrac{1}{2}\phi +
 \udot}\lc_{ab}\El^b =
4\Psi_{3a}
 +\phi\chi_{1a} + 2\chi_{2a}.\label{E_ahat}
\ea
\end{mathletters}
The terms on the left are those which govern an EM field around a BH; those on
the right are the interaction terms. Note that these equations are a mixture of
first and second order quantities, and are thus not gauge-invariant, and
therefore not integrable.

In order to convert ME into gauge-invariant form, it is not enough to define
the interaction variables above; we must also do something with the magnetic
field: in ${\cal S}$ the magnetic field appearing in ME has a contribution
from the static background field in ${\cal F}_1$ which we must somehow
subtract off. The standard route to do this is as a series expansion, but
this does not work here. {If we imagine that $B_a$ is written as a power
series,
\be
B^a=\epsilon_\B \bra{B_1^a+\epsilon_g B_2^a +\cdots}
\ee
where $B_1^a$ satisfies the ${\cal F}_1$ equations,~(\ref{B1eqs}), then one
would imagine $B_1^a$ would cancel out of the ${\cal S}$ ME when $B^a$ appears
alone leaving just $B_2^a$; when it appears multiplying an ${\cal F}_2$ term it
is only $B_1^a$ which contributes. However, this is not the case. It is
possible to show from the commutation relations for the hat and dot derivatives
acting on $B^a$ that this leads to an inconsistency, implying that the
interaction terms must be zero. Consider, for example, the scalar part of the
magnetic field:
\be
\B=\epsilon_\B \B_1+\epsilon_g\epsilon_\B \B_2+{\cal
  O}(\epsilon_g^2,\epsilon_\B^2)\label{Bexpans}
\ee
where $\B_1$ satisfies $\dot{\B}_1=0$ and $\hat{\B}_1=F$ where $\dot{F}=0$,
representing the background solution ${\cal F}_1$. Now, using the commutation
relation given by Eq. (30) in~\citet{CB},
\be
\hat{\dot{\B}} = \epsilon_g\epsilon_\B\hat{\dot{\B}}_2 =
\epsilon_g\epsilon_\B\bra{\dot{\hat{\B}}_2-\udot\dot{\B}_2},
\ee
by using the commutation relation after substituting from Eq~(\ref{Bexpans}),
and neglecting terms ${\cal O}(\epsilon_g^2)$. Alternatively,
\ba
\hat{\dot{\B}} = \dot{\hat{\B}} - \udot\dot{\B} - 2\alpha_a\delta^a\B
= \epsilon_g\epsilon_\B\bra{\dot{\hat{\B}}_2 - \udot\dot{\B}_2} -
2\epsilon_\B\alpha_a\delta^a\B_1 ,
\ea
where we applied the commutator \emph{before} using the expansion given by
Eq.~(\ref{Bexpans}). This is clearly a contradiction if
$\alpha_a\delta^a\B_1\neq0$, which is the case here. [The correct form of
calculating this equation results in Eq.~(\ref{hatbeta}).] In fact, this
problem usually arises when using covariant (partial-)frame methods for
second order perturbation theory. In contrast to metric-based approaches, the
solutions for perturbed derivative \emph{operators} are never sought, so
\emph{they must always operate on quantities of the same perturbative order}.
We must therefore define some gauge-invariant variables for the magnetic
field.

\subsubsection{Gauge-invariant variables for the magnetic
field}

We define the variables
\be
\beta\equiv\dot\B,~~~~\mbox{and}~~~~\beta_a\equiv\dot\B_{\bar a}
=\dot\B_a+\bra{\alpha^b\B_b}\n_a,\label{dotB_adef}
\ee
which are gauge-invariant in ${\cal S}$, as they vanish in ${\cal
F}$~\citep{BS,Bruni-Sonego99}. To convert ME into a gauge-invariant
system of equations,  we
must somehow replace every occurrence of $\B$ with $\beta$, and $\B_a$ with
$\beta_a$. Note first that
\be
\beta=-\lc_{ab}\delta^a\El^b,\label{beta}
\ee
immediately from Eq.~(\ref{BSdot}). Meanwhile, the commutation relation between
hat- and dot-derivatives (see Eq. (30) in~\citet{CB}), when applied to $\B$
results in the propagation equation
\be
\hat{\beta} = -\bra{\phi + \udot}\beta - \delta_a\beta^a +
4\delta_a\Psi_3^a + \phi\delta_a\chi_1^a
+ 2\delta_a\chi_2^a,\label{hatbeta}
\ee
where we have used Eq.~(\ref{hatB2}), and the appropriate commutation
relation for dot-$\delta$ derivatives on vectors. However, this equation also
arises from propagating~(\ref{beta}) using ME, as it should. Hence, because
Eq~(\ref{beta}) is a consistent constraint, the propagation equation for
$\beta$ is redundant. This implies that Eq.~(\ref{beta}) can replace
Eqs.~(\ref{hatB2}) and~(\ref{BSdot}). To find a propagation equation for
$\beta_a$ we must propagate Eq.~(\ref{dotB_adef}) using the appropriate
commutation relation for vectors, giving
\ba
\hat{\beta}_a = \lc_{ab}\ddot{\El}^b - \bra{\tfrac{1}{2}\phi + 2\udot}\beta_a
+ \delta_a\beta
 - 2\ddot{\chi}_{1a} - \phi\chi_{3a} - 2\chi_{4a}
-
 2r^{-2}\Psi_{1a}-4r^{-2}M_{1a} - 2r^{-2}J_{1a}, \label{hatbeta_a}
\ea
which replaces Eq.~(\ref{Bahat}). It is this equation which brings Weyl
curvature into ME through the commutation relations. A key remaining
evolution equation comes from calculating $\ddot
\El$ using Eq.~(\ref{Esdot}):
\be
\ddot{\El} = \lc_{ab}\delta^a\beta^b + \phi\lc_{ab}\delta^a\chi_1^b +
2\lc_{ab}\delta^a\chi_2^b
,\label{EScons}
\ee
which propagates consistently. We will use this evolution equation, which is
just the gauge-invariant form of Eq~(\ref{Esdot}), to replace
Eq.~(\ref{Eshat}). Therefore, \emph{ME are now just the two vector
propagation equations~(\ref{hatbeta_a}) and~(\ref{E_ahat}), together with the
two scalar non-propagation equations~(\ref{beta}) and~(\ref{EScons})}. The
last two serve as definitions for $\beta$ and $\El$ after time harmonics are
used; these then become constraints.

Note how converting the gauge-dependent form of ME,
Eqs.~(\ref{Eshat})~--~(\ref{E_ahat}), which contain a mixture of first and
second perturbation orders, into a gauge-invariant \emph{second order} system
has introduced many more interaction terms into the equations, terms arising
purely from the Ricci identities. These terms are essentially hidden in the
frame derivatives (dot, hat and $\delta$) when acting on $\B$ in
Eqs.~(\ref{Eshat})~--~(\ref{E_ahat}), and in $\B$ itself, illustrating the
importance of using a full set of gauge-invariant variables.

Although equations we have derived are gauge-invariant to order ${\cal
O}\bra{\epsilon_\B\epsilon_g}$, they are actually valid up to ${\cal
O}\bra{\epsilon_\B^2\epsilon_g}$, ${\cal O}\bra{\epsilon_\B^3}$, which can be
easily seen as follows. If we include terms ${\cal O}\bra{\epsilon_\B^2}$
(i.e., the energy density and anisotropic pressure of the static magnetic
field) in the gravity sector, then changes to $W_{ab}$ (and other GW variables)
is ${\cal O}\bra{\epsilon_\B^2}$, making the change to the GW-B variables
${\cal O}\bra{\epsilon_\B[\epsilon_g+\epsilon_\B^2]}$. However, the equations
are not gauge-invariant at this order, because the variables are non-zero at
lower perturbative order (i.e., ${\cal O}\bra{\epsilon_\B\epsilon_g}$); see,
e.g.,~\citet{BS,Bruni-Sonego99}.

The gauge-invariant form of the equations now shows exactly the terms and
couplings involved in generating the EM field. Consider, for example, the
covariant wave equation for $\El$:
\ba
-\ddot{\El} + \hat{\hat{\El}} + \delta^2\El + \bra{2\phi +
\udot}\hat{\El} - \bra{\tfrac{1}{2}\phi^2+2\E}\El
 = -2\phi\lc_{ab}\delta^a\chi_1^b - 4\lc_{ab}\delta^a\chi_2^b
 - 4\lc_{ab}\delta^a\Psi_3^b.
\ea
The lhs of this equation is just the contribution from the BH geometry, and
can be related simply by a change of variables to the usual Regge-Wheeler
equation for an electromagnetic field around a BH [compare with
Eq.~(\ref{RW})~-- see also Eq.~(\ref{wave1})]. The rhs, on the other hand is
the source from the interaction terms, and has contributions from the time
integral and angular derivative of the dot-product between the
transverse traceless
RW shearing tensor and the angular (sheet) part of the magnetic
field, and the 2-divergence of the RW tensor times the magnitude
of the radial part of the magnetic field.

\subsection{The initial value  and quasi-normal mode formulations}

\noindent\textsc{spherical harmonics}:~
In order to numerically integrate the system of equations we must split them
using spherical harmonics, which removes the tensorial nature of the equations,
and turns Eqs.~(\ref{GWBcons}) into \emph{algebraic} relations. In the Appendix
we have given an overview of the spherical harmonics we use, which were
developed in~\citet{CB}.

A spherical harmonic decomposition of all variables then implies, from
Eqs.~(\ref{GWBcons}), that the spherical harmonic components of each of the
variables $\bm K^\lBg,$ $\bm
\Psi_a^\lBg$, $\bm M_a^\lBg$, $\bm J_a^\lBg$ are proportional to the harmonic
components of $\bm\chi_a^\lBg$. So, for example, for each $\ell$:
\begin{mathletters}
\ba
\bm\Psi\V^\lBg&=&\frac{\capl_\B\capl_g}{\capl_\B-\capl}\bm\chi\V^\lBg,\\
\bar{\bm\Psi}\V^\lBg&=&\frac{\capl_\B\minl_g\bra{\capl+\capl_g-\capl_\B}}{\bra{\capl_\B+\capl}\capl_g-\bra{\capl-\capl_\B}^2}
\bar{\bm\chi}\V^\lBg,
\ea
\end{mathletters}
with similar relations for $\bm K^\lBg,\bm M_a^\lBg,\bm J_a^\lBg$.

Because the equations are linear in the second order variables, when we split
into spherical harmonics, the equations decouple into two distinct subsets of
opposing parity; the \emph{parity mixing} which occurs between the magnetic
field and the GW is contained in the interaction variables. We will call the
set of equations containing $\El\V$ the \emph{even parity equations}, and
those containing $\bar\El\V$ the \emph{odd parity equations}. Unfortunately,
all the other variables are of the `opposite' parity to $E_a$ in each system
of equations, so this may cause confusion (so, e.g., $\bar\beta\V$ and
$\bar{\bm\chi}\V$ are of even parity, etc.).

\vskip\baselineskip
\noindent\textsc{initial value formulation}:~
A useful form of writing ME is as  wave equations. For this we will use the
variables
\begin{mathletters}
\ba
W\sfS&=&r^{2}\El\sfS~~~~~~\mbox{(\sc even)},\\
\bar W\sfS&=&\tfrac{1}{2}\phi r^3\beta\sfS~~~\mbox{(\sc odd)}.
\ea
\end{mathletters}
Then these variables satisfy the wave equations for each $\ell$:
\be
-\ddot W+\hat{\hat W}+\udot\hat W{-\frac{\capl}{r^2}}W=S\bra{\bm
\chi_a}\label{wave1}
\ee
where $W=\brac{W\sfS,\bar W\sfS}$ and we have defined the even and odd source
terms as
\begin{mathletters}
\ba
S\sfS &=& -2\capl r \sum_{\ell_\B=1}^\infty\sum_{\ell_g=2}^\infty \brac{ \phi
\bar\chi_{1\V}^\lBg + 2\bar\chi_{2\V}^\lBg
+2\frac{\minl_g\bra{\capl + \capl_g - \capl_\B}}{{\bra{\capl_\B-\capl}^2 -
    \bra{\capl_\B+\capl}\capl_g}}\bar\chi_{3\V}^\lBg}, \\
\bar S\sfS &=& \capl
\sum_{\ell_\B=1}^\infty\sum_{\ell_g=2}^\infty\Bigg\{-2\phi
r^2\ddot\chi_{1\V}^\lBg\!\!\!
-2\phi\frac{\minl_g\bra{\capl-3\capl_\B}}{\capl-\capl_\B}\chi_{1\V}^\lBg
\nonumber \\
&&\qquad\qquad\quad  +\frac{\bra{\capl_\B-4\minl_g-\capl}\phi^2r^2+4\minl_g}{\capl-\capl_\B}\chi_{3\V}^\lBg
\vphantom{\Bigg(}
+2r^2\frac{\capl_\B+\minl_g-\capl}{\capl-\capl_\B}\chi_{4\V}^\lBg
\Bigg\}\label{waveend}.
\ea
\end{mathletters}
These wave equations may replace
ME, and, more importantly, are in the form of an initial value problem. We
then have, for each parity, one forced wave equation for the EM field, plus
two evolution equations for the interaction variables ($\chi_1$ and
$\chi_3$), plus two constraints (propagation equations); the set of four DEs
for the interaction variables may be easily turned into a set of two coupled
wave equations instead by eliminating two $\bm\chi_a$ variables (either
$\chi_1$ and $\chi_3$ or $\chi_2$ and $\chi_4$). Eliminating $\chi_{2a}$
using the first equation of~(\ref{chiprop}) and $\chi_{4a}$ using the third
equation turns the remaining two into wave equations: for the odd parity
equations, we find
\begin{mathletters}
\ba
&&-\ddot{\chi}_{1\V}^\lBg + \hat{\hat{\chi}}_{1\V}^\lBg +
\udot\hat{\chi}_{1\V}^\lBg =
- 2\bra{\phi + \udot}\hat{\chi}_{1\V}^\lBg +
2\hat{\chi}_{3\V}^\lBg\nonumber\\
&&\qquad\qquad +\bras{6\E - \tfrac{1}{2}\phi^2 - \udot^2 +
\bra{\capl_g
-\capl_B}r^{-2}}\chi_{1\V}^\lBg + \bra{3\phi + 2\udot}\chi_{3\V}^\lBg,\\
&&-\ddot{\chi}_{3\V}^\lBg + \hat{\hat{\chi}}_{3\V}^\lBg +
\udot\hat{\chi}_{3\V}^\lBg = 2\capl_\B r^{-2}\hat{\chi}_{1\V}^\lBg
-{4\phi}\hat{\chi}_{3\V}^\lBg\nonumber\\
&&\qquad\qquad + 2\capl_\B r^{-2}\bra{\phi + \udot}\chi_{1\V}^\lBg
+ \bras{7\E - 3\phi^2 +
\bra{\capl_g-\capl_B}r^{-2}}\chi_{3\V}^\lBg, 
\ea\label{sources}
\end{mathletters}
with identical equations for the even variables.

The full solution for the induced EM radiation is given by the variables $W$:
for \emph{even} perturbations, $\El\V$ is given by~(\ref{Eshat}) and
$\bar\beta\V$ by~(\ref{EScons}); for \emph{odd} perturbations, $\bar\El\V$
by~(\ref{beta}) and $\beta\V$ by~(\ref{hatbeta}).

\vskip\baselineskip
\noindent\textsc{quasi-normal mode formulation using temporal harmonics}:~
While the covariant equations above are given with time derivatives, allowing
the problem to be put in the form suitable for solving as an initial value
problem, it is often advantageous to use time harmonics. In particular, the
effect of BH ringdown is conventionally studied by this method, as the
ringdown phase is characterised by a set of quasi-normal
frequencies~\citep{QNM,Kokkotas-Schmidt99}, which are independent of
the initial perturbation. We
achieve this by replacing all dot-derivatives by a factor of $i\omega$, with
the usual understanding that subsequent equations are then for the spatial
parts only~\citep{CB}, although formally it is significantly more
complicated~\citep{QNM,Kokkotas-Schmidt99,andersson2}. The harmonic
function $\omega$ is defined
with respect to the proper time, $\tau$, of observers travelling on $u^a$,
and satisfies Eq.~(\ref{omdef}); $\sigma$ is the \emph{constant} harmonic
index associated with time, $t$, measured by observers at infinity. Note that
they are related by $\omega\tau=\sigma t$.

The time derivative of the second order interaction variables $\bm
K^\lBg,\bm\Psi_a^\lBg,\bm\chi_a^\lBg,\bm M_a^\lBg,\bm J_a^\lBg$
acts only on the GW part of the term,
because the time derivative of the magnetic field is already second order.
Therefore, when these terms are split into time harmonics, and the
interaction equations~(\ref{GWBprop}) are solved, the usual boundary
conditions on the GW variable $W_{ab}$ will take effect~-- that the GW cannot
propagate out of the horizon, or in from infinity. This implies that the
allowed frequencies $\sigma$ must be discrete with positive imaginary
part~\citep{QNM,Kokkotas-Schmidt99}. This represents modes which decay
exponentially in time, but
whose amplitudes grow exponentially with radius.

Our method presented here, which sets up the equations as a set of
purely second order, linear, gauge-invariant differential
equations, means that when we solve them we don't view quadratic
first order effects as quadratic forcing terms in the second order
equations, but as second order quantities in their own right; the
first order equations are forgotten about. Therefore the
propagation equations governing the interaction variables,
Eqs.~(\ref{chiprop}) and~(\ref{GWBprop}), also must be confined to
these frequencies. Hence, the coupling between the equations for
the induced EM field and the interaction variables implies that
the \emph{allowed independent frequencies of the induced EM
radiation must be identical to those of the forcing GW}~-- the
quasi-normal frequencies; that is, the GW and EM radiation satisfy
the same dispersion relation, and are in resonant interaction.
Other frequencies correspond to EM waves which are not induced by
the interaction terms with these boundary conditions (and form
part of the homogeneous solution for the EM field); there is no
need to consider these here. Therefore, when we split the system
of equations using the time harmonics, each $\ell_g$ picks out a
set of allowed frequencies in the interaction equations, thus
\emph{removing the summations over $\ell_g$} in ME. For each
$\ell_g$ there is one system of equations for each quasi-normal
frequency~$\omega_\elg$ associated with that particular $\ell_g$.
The complete solution for $\El\V$, for example, may then be
written schematically \emph{for each $\ell$} as
\ba
\El\V=\sum_{\ell_g=2}^\infty \El\V\bra{\ell_g}=
\sum_{\ell_g=2}^\infty \sum_{~~{\omega \in \brac{\omega_g}_n}}
\El\V^{(\omega)}\bra{\ell_g}e^{i\omega\tau}
\ea
where $\brac{\omega_g}_n$ denotes the set of all quasi-normal frequencies for
a given $\ell_g$.

From the wave equations give above, it is clear that for each parity, while
there are three EM variables, there are only two degrees of freedom in the EM
radiation; in the even case for example these are $W\sfS$ and $\hat W\sfS$,
resulting in a straightforward wave equation. We can of course stick to these
variables in the QNM formulations of the problem, but the system is naturally
first order in the variables $\El\V$ and $\bar\beta\V$ (or $\El\sfS$) in the
even case once the extra degree of freedom is removed (similarly for the odd
case). There doesn't seem much advantage whichever way we choose the
variables so we will remove the scalar (radial) parts of the EM field from
the system of equations, using Eqs.~(\ref{beta})
and~(\ref{EScons}). Our key equations then become:
\begin{mathletters}
\ba
&&\!\!\!\!\!\!\!\!\!\!\!\!\!\!\!\mbox{\sc even parity}:\nonumber\\
&&\!\!\!\!\!\!\!\!\!\hat{\El}\V =
-\bra{\tfrac{1}{2}\phi + \udot}\El\V + \bra{1 -
{\capl}{\omega^{-2}r^{-2}}}\bar\beta\V\nonumber\\
    && + \sum_{\ell_\B = 1}^{\infty}\brac{-\bra{1 + \capl\omega^{-2}r^{-2}}
    \bras{\phi\bar{\chi}_{1\V}^\lBg + 2\bar{\chi}_{2\V}^\lBg}
    + 4\frac{\minl_g\bra{\capl + \capl_g - \capl_\B}}{{\bra{\capl_\B -
      \capl}^2 -
    \bra{\capl_\B+\capl}\capl_g}} \bar{\chi}_{3\V}^\lBg}  , ~~~~~~~~
    \label{MEGI1}       \\
&&\!\!\!\!\!\!\!\!\!\hat{\bar\beta}\V = -\bra{\tfrac{1}{2}\phi +
     2\udot}\bar{\beta}\V -
     \omega^2\El\V 
    + \sum_{\ell_\B = 1}^{\infty}
    \bras{2\bra{\omega^2 - \minl_gr^{-2}}\bar{\chi}_{1\V}^\lBg -
    \phi\bar{\chi}_{3\V}^\lBg
    -2\bar{\chi}_{4\V}^\lBg}   ,  \label{MEGI2} \\
&&\!\!\!\!\!\!\!\!\! \hat{\bar{\bm\chi}}^\lBg\V =
    \bm\Gamma^\lBg{\bar{\bm\chi}}^\lBg\V;\\
&&\!\!\!\!\!\!\!\!\!\!\!\!\!\!\!\mbox{\sc odd parity}:\nonumber\\
&&\!\!\!\!\!\!\!\!\!\hat{\bar{\El}}\V =
-\bra{\tfrac{1}{2}\phi + \udot}\bar{\El}\V -
    \beta\V
+\sum_{\ell_\B = 1}^{\infty}\brac{
    \bras{\phi{\chi}_{1\V}^\lBg + 2{\chi}_{2\V}^\lBg}
    + 4\frac{\minl_g}{\capl_\B - \capl}{\chi}_{3\V}^\lBg}        ,  \label{MEGI3}  \\
&&\!\!\!\!\!\!\!\!\! \hat{\beta}\V = -\bra{\tfrac{1}{2}\phi + 2\udot}\beta\V -
    \bra{-\omega^2 +
  \capl r^{-2}}\bar{\El}\V\nonumber\\&&
    + \sum_{\ell_\B = 1}^{\infty} \brac{2\bras{\omega^2 + \frac{\bra{\capl -
    3\capl_\B}\minl_g}{\bra{\capl_\B - \capl}r^2}}{\chi}_{1\V}^\lBg
    - \phi{\chi}_{3\V}^\lBg - 2{\chi}_{4\V}^\lBg},\label{MEGI4}\\
&&\!\!\!\!\!\!\!\!\!\hat{\bm\chi}^\lBg\V =
    \bm\Gamma^\lBg\bm\chi^\lBg\V,
\ea\label{alleqnsevenodd}
\end{mathletters}
for each $\ell_g$ and  $\omega \in \brac{\omega_g}_n$. Each parity
consists of a
set of six coupled ordinary differential equations in the radial parameter
$\rho$.

\forget{
\ba EQUATIONS IN TERMS OF PSI
\mbox{\sc even parity}:\nonumber\\
\hat\El\V&=&-\bra{\tfrac{1}{2}\phi+\udot}\El\V+\bra{1-{\capl}{\omega^{-2}r^{-2}}}\bar\B\V\nonumber\\
    &&+\sum_{\ell_\B}\brac{\bra{1+\capl\omega^{-2}r^{-2}}
    \frac{{\bra{\capl_\B-\capl}^2-\bra{\capl_\B+\capl}\capl_g}}
    {\minl_g\capl_\B\bra{\capl+\capl_g-\capl_\B}}\bras{\phi\bar{\Psi}_{1\V}^\lBg+2\bar{\Psi}_{2\V}^\lBg}
    -4\capl_\B^{-1}\bar{\Psi}_{3\V}^\lBg}  ,         \\
\hat{\bar\B}\V&=& -\bra{\tfrac{1}{2}\phi+2\udot}\bar\B\V -\omega^2\El\V\nonumber\\&&
    +\sum_{\ell_\B}\frac{{\bra{\capl_\B-\capl}^2-\bra{\capl_\B+\capl}\capl_g}}
    {\minl_g\capl_\B\bra{\capl+\capl_g-\capl_\B}}
    \bras{2\bra{-\omega^2+\minl_gr^{-2}}\bar{\Psi}_{1\V}^\lBg +\phi\bar{\Psi}_{3\V}^\lBg
    +2\bar{\Psi}_{4\V}^\lBg  }   ,   \\
\hat{\bar{\b\Psi}}^\lBg_a &=&\b\Gamma^\lBg{\bar{\b\Psi}}^\lBg_a;\\
\mbox{\sc odd parity}:\nonumber\\
\hat{\bar\El}\V&=& -\bra{\tfrac{1}{2}\phi+\udot}\bar\El\V -\B\V\nonumber\\&&
    +\sum_{\ell_\B}\brac{\frac{\capl_\B-\capl}{\minl_g\capl_\B}
    \bras{\phi{\Psi}_{1\V}^\lBg+2{\Psi}_{2\V}^\lBg}
    -4\capl_\B^{-1}{\Psi}_{3\V}^\lBg}        ,    \\
\hat\B\V&=& -\bra{\tfrac{1}{2}\phi+2\udot}\B\V -\bra{-\omega^2+\capl r^{-2}}\bar\El\V\nonumber\\&&
    +\sum_{\ell_\B} \frac{\capl_\B-\capl}
    {\minl_g\capl_\B}\brac{2\bras{\omega^2+\frac{\bra{\capl-3\capl_\B}\minl_g}{\bra{\capl_\B-\capl}r^2}}{\Psi}_{1\V}^\lBg
    -\phi{\Psi}_{3\V}^\lBg-2{\Psi}_{4\V}^\lBg},\\
\hat\b\Psi^\lBg\V &=&\b\Gamma^\lBg\b\Psi^\lBg\V,
\ea
}

\section{Numerical examples}\label{sec4}

We have now set up the equations as a gauge-invariant linear system of
differential equations in purely second order variables, in two different
ways. The first is as a set of three coupled wave equations (for each
parity), which may be numerically integrated as an initial value problem once
some initial data is specified. The second is as a six-dimensional system of
first-order ordinary differential equations (for each parity) which are
Fourier decomposed in time, which is suitable for integration once
appropriate boundary conditions are satisfied. There are of course advantages
and disadvantages to both, which we discuss presently.

While it would be desirable to be able to integrate these equations in a
situation which is astrophysically accurate in some sense, this is quite a
non-trivial problem as it involves specifying initial data from a fully
non-linear integration of the field equations in a situation such as, for
example, BH-BH merger. This is beyond the purpose of the present discussion, as
we would like to get an overall estimate of the strength and importance of the
effect in this first instance.

In general, the summations over $\ell_g$ and $\ell_\B$ in the equations for the
generated EM radiation mean that these coupled systems of equations are
infinite dimensional. However, for a static magnetic field around a BH the
dominant contribution to the field strength will be dipolar, and the GW emitted
by a compact object will typically be dominated by the quadrupole radiation
(for example, when two BHs collide head-on from an initially small separation,
the emitted radiation is pure quadrupole~\citep{PP}; other studies with high
energy collisions support this conclusion~\citep{card,card2,card3,card4}).
Therefore, in this section we will investigate numerically the
$\ell_g=2,~\ell_\B=1$ interaction while ignoring the contribution from the
others.

As we mentioned earlier, the case of an $\ell_\B=1$ magnetic field has two
solutions, one which is uniform at infinity, and one which falls off at large
distances  like $1/r^3$, a dipole. Both of these are of interest
astrophysically, as magnetic fields surrounding compact objects can extend
considerable distances when supported by a plasma (i.e., a BH `embedded' in an
external magnetic field), but be purely dipolar close in. It is clearly
important to distinguish the two cases in the $\bm\chi_a$ variables when we
integrate the equations, in order to determine which type of field is
responsible for what. For both solutions, the ratio $\B\V/\B\sfS$ is a known
function of $r$, with no dependence on any boundary conditions~-- see, e.g.,
Eq.~(\ref{Bdipolesol}). This implies that
\be
\frac{\chi_1}{\chi_3}=\frac{\chi_2}{\chi_4}=r\left.\frac{\B\V}{\B\sfS}\right|_{\mbox{uniform
\emph{or} dipole solution}},\label{unidip}
\ee
where $\chi_i$ ($i=1,\cdots,4$) represents either the odd or even
parity part of $\chi_{ia}$. The ratio $\B\V/\B\sfS$ is given by
Eq.~(\ref{Bdipolesol}) for the dipolar field, while for the field
which is uniform at infinity (characterised by $\hat\B\sfS=0$) it
equals $\tfrac{1}{2}\phi r$. Thus, if we desire the magnetic field
to be one of these solutions, we can use Eq.~(\ref{unidip}) to
constrain the boundary conditions, or simply replace
$\chi_3,~\chi_4$ in the equations. In Fig.~\ref{chi1chi3} we show
a plot of the ratio $-\chi_3/\chi_1$ for the pure dipole field
which shows how the dominant contribution to the interaction terms
at large distances and close to the horizon is dominated by
$\chi_1$ (or $\chi_2$; the figure for $-\chi_2/\chi_4$ is
identical). Thus, $\chi_3$ and $\chi_4$, containing the radial
part of the magnetic field, only contribute significantly in the
vicinity of the photon sphere. We will consider only the pure
dipole solution here, and hereafter remove $\chi_3,\chi_4$ using
Eq.~(\ref{unidip}) (we remove these two because, as
Fig.~\ref{chi1chi3} shows, replacing $\chi_1,\chi_2$ would make
numerical solutions become unstable at small and large distances).
\begin{figure}[ht!]
\begin{center}
\includegraphics[width=0.7\textwidth]%
{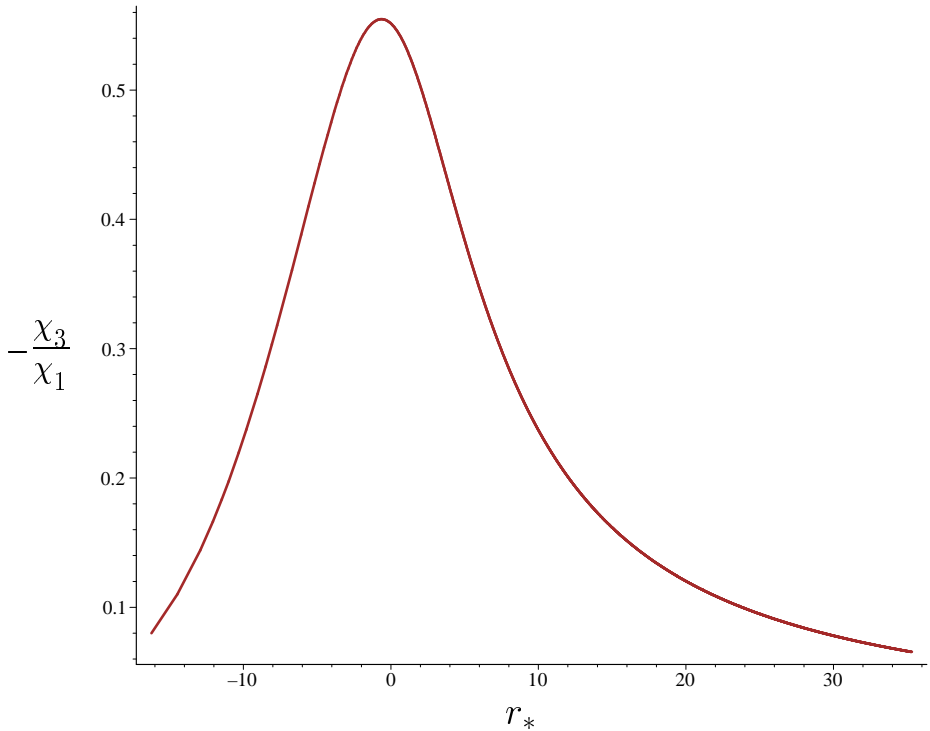}
\end{center}
\caption{For a pure dipole background magnetic field this
figure, which is a plot of Eq.~(\ref{unidip}), shows how the relative
contributions from the interaction terms are dominated by $\chi_1$ (or
$\chi_2$), except in the region just inside the photon sphere (the peak is at
about $r\sim2.4m$), where $\chi_3$ becomes significant (recall that $\chi_3$ is
defined using the angular gradients of the radial part of the background
magnetic field). The `tortoise' coordinate $r_*$ is defined by
Eq.~(\ref{tortoise}) below.
\label{chi1chi3}}
\end{figure}

The induced EM radiation will of course be of much higher amplitude far from
the BH if we allow for the presence of the uniform magnetic field as part of
the static background, as the  interaction distance will be vastly increased.
For the pure dipole magnetic field, the interaction distance is effectively
curtailed at large $r$, because the magnetic field strength falls off so fast
that $\El\gg\chi$ by $r\sim20m$ or so. In astrophysical situations where the
magnetic field extends far from the source (supported by an accretion disk,
or entangled in the ejected envelope of the progenitor star, for example), we
would expect further, linearly growing amplification, (beyond say $r\sim
20m$) over the amplification we report below. This will be studied at a later
date when a plasma is included into the discussion, but should be borne in
mind in what follows.

Hereafter we shall set $m=1$ (which just defines the units of $r$), and we
shall use the tortoise coordinate $r_*$ of Regge and Wheeler~\citep{RW} defined
by
\be
{d\rho}=\tfrac{1}{2}\phi r{dr_*}=\bra{\tfrac{1}{2}\phi
r}^{-1}dr~~~\Rightarrow~~~r_*=r+2m\ln\bra{\frac{r}{2m}-1}.\label{tortoise}
\ee
Because the system of equations we are investigating
are linear, the units we use are physically irrelevant, and is tied into the
physical amplitude of our initial data which we normalise at unity (so that if
units are chosen for $\chi$ say we can immediately read off the actual
amplitudes for $\El$).

\subsection{The initial value problem}

Here we envisage the following situation: at some initial time $t=0$ the
interaction is `turned on' with some typical initial profile for the GW
[i.e., the tensor $W_{ab}$, which translates in this case to
$\bm\chi^a(t=0)=\bm\chi^a_0$], at which time the induced EM field is zero, but
with non-zero second time derivatives (`acceleration'). Although intuitively
reasonable for modeling a situation such as BH formation or where the
magnetic field becomes very strong very quickly, say, we require this
switching on of the interaction because otherwise the ME will not be
consistent for a general $\bm\chi^a_0$.

A common way of specifying initial data for this type of problem is to
consider GW scattering off a BH, with the initial data given by a static
narrow Gaussian peak at some distance from the hole~\citep{andersson2}. This
then splits in two as the RW equation is evolved, with the part falling into
the hole of most interest: this scatters off the photon sphere and starts the
black hole vibrating (roughly speaking), with a characteristic waveform which
is largely independent of the initial data, dominated by the quasi-normal
modes of the BH (which only depend on its
mass)~\citep{andersson1,andersson2,SP,QNM}. We will use this scenario with
$W_0\sim\exp{\bra{-(r_*-20)^2}}$ at $t=0$, which we normalise so that at
$t=0$ and $r_*=20$, $\chi_{1\V}=1$. We will not consider a pulse originating
further from the hole because the dipole field falls off so fast with
distance; the qualitative results remain the same.

We  then evolve our key equation~(\ref{wave1}) and the wave equation for
$\chi_1$ [modified by replacing $\chi_3$ with Eq.~(\ref{unidip}) as discussed
above] with this initial data. This then gives the solution for $W\sfS$, which
we convert to $\El\V$. Results are shown in Figs.~(\ref{IVP3d})
and~(\ref{IVP2}) for $\log_{10}|\El\V|$.
\begin{figure}[ht!]
\begin{center}
\includegraphics[width=0.7\textwidth]{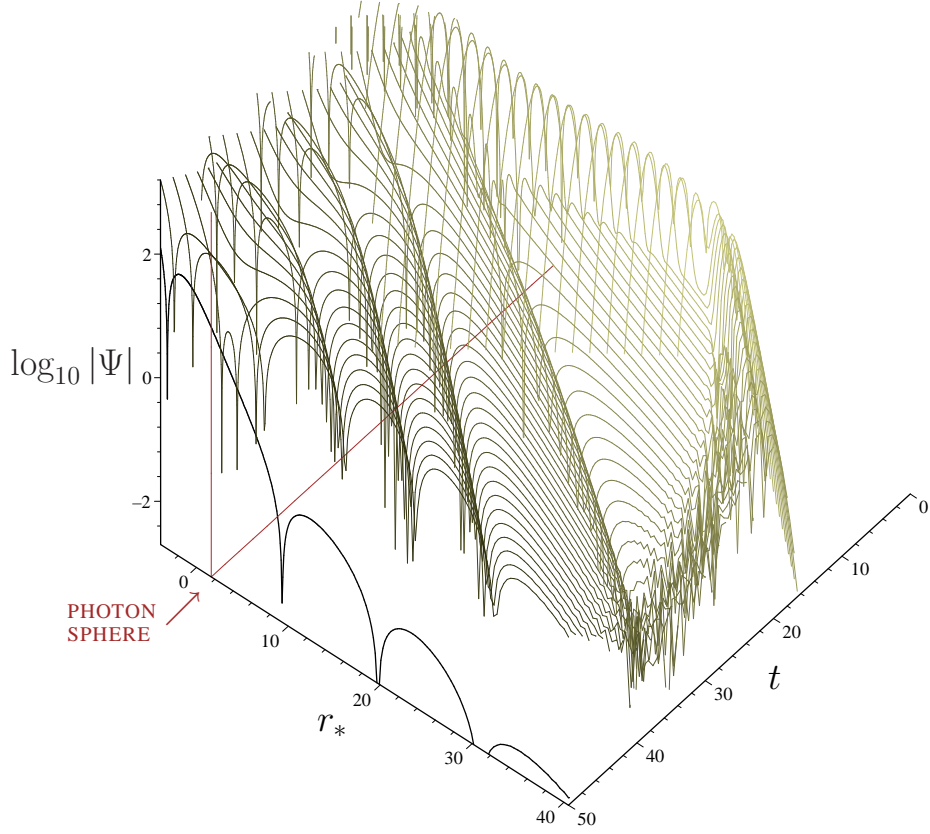}
\end{center}
\caption{
The induced EM radiation from static Gaussian initial data for the GW at
$t=0$, such that the EM field is zero. This pulse splits into two, one
falling into the hole and the other propagating to infinity. The part falling
into the hole is partially reflected at about $t\sim15$ generating the
`ringing' we see later at $t=50$ in the thick curve (modulated by the
magnetic field here~-- this is $\chi_{1\V}$). During this in-fall, the GW-B
interaction produces substantial amounts of EM radiation which is reflected
back away from the hole, and is further increased by the subsequent BH
ringing.
\label{IVP3d}}
\end{figure}
\begin{figure}[ht!]
\begin{center}
\includegraphics[width=0.7\textwidth]{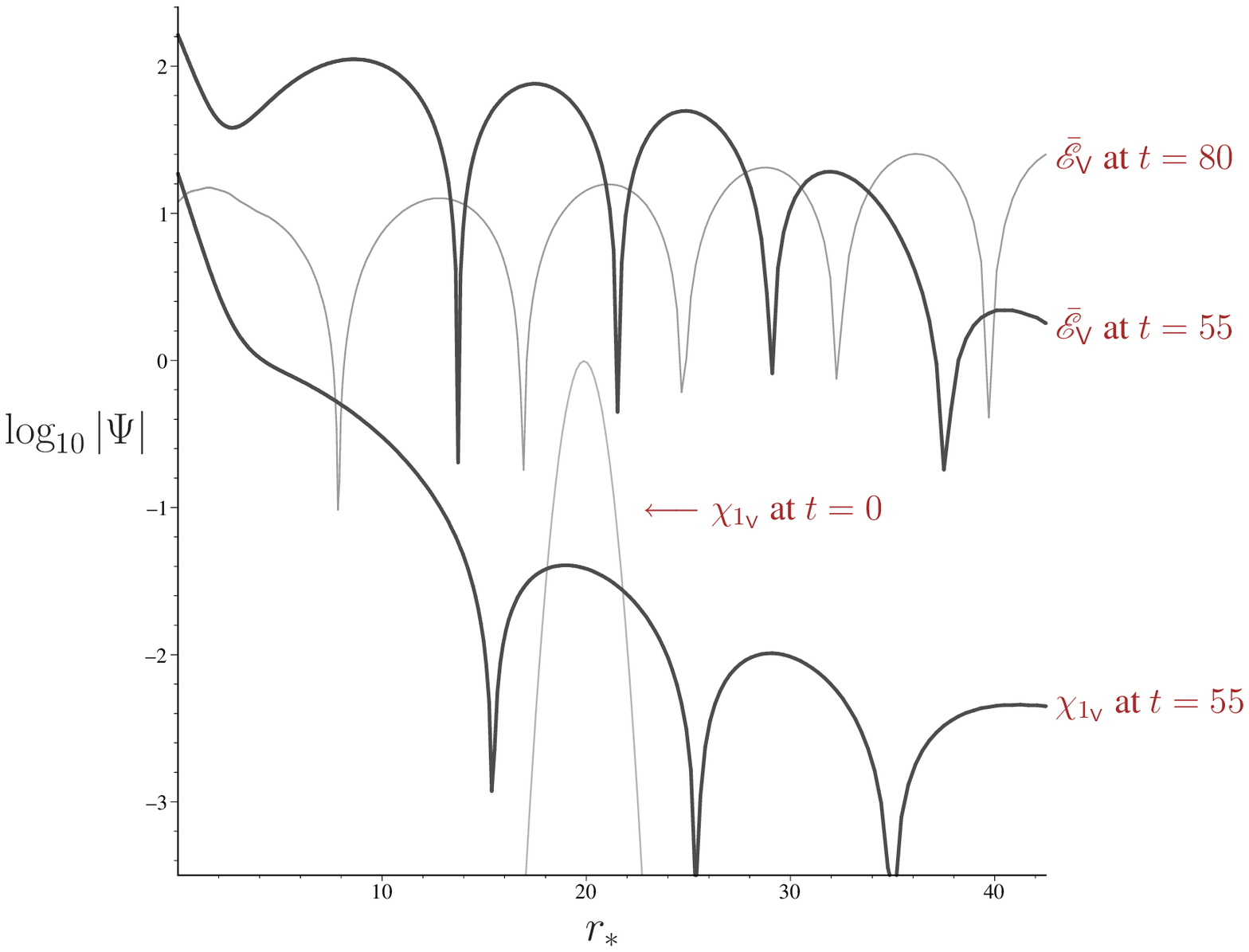}
\end{center}
\caption{The induced EM radiation from static Gaussian initial data
  for the GW at
$t=0$, such that $W\sim\exp{\bra{-(r_*-20)^2}}$, normalised so
that at $t=0$ and $r_*=20$, $\chi_{1\V}=1$; at $t=0$ the EM field
is zero. This pulse splits into two, one falling into the hole and
the other propagating to infinity. The part falling into the hole
is partially reflected generating the `ringing' we see at $t=55$
(modulated by the magnetic field here in $\chi_{1\V}$), in the
wave which is moving to the right . During this in-fall, the GW-B
interaction produces substantial amounts of EM radiation, which is
further increased by the subsequent BH ringing. Thus we see at
$t=55$ the EM field generated is two orders of magnitude larger
than the initial pulse, and over three orders of magnitude larger
than the interaction terms at the same distance from the hole. At
$t=80$ we see that the main amplification has taken place with the
peak of the  induced EM radiation moving past
$r_*\sim40$.\label{IVP2}}
\end{figure}
These figures show the EM radiation generated and subsequently amplified
during the scattering of the GW off the photon sphere. The ringing of the BH
then generates a continuous stream of EM radiation, which at its peak is over
\emph{two orders of magnitude} larger than the initial pulse of radiation (by
the time it is reflected back out to $r_*=20$). This radiation mirrors very
closely the GW waveform making it a suitable EM counterpart for GW emission.

\subsection{The quasi-normal mode approach}

We shall now integrate the equations in the frequency domain, summing over the
QNMs of the BH, which will tell us about the strength of the interaction in the
latter stages of a perturbation of a BH independently of the initial
perturbation~\citep{andersson2}. We imagine that the interaction starts at
$t=0$ at some inner radius $r_0$, so for $r<r_0$ we assume that
$\El\sfS=\beta\sfS=0$, while $\bm\chi$ does its own thing; at $r=r_0$ we choose
boundary conditions for each $\omega_n$ such that all EM terms and their
derivatives are equal to zero; for want of accurate boundary conditions for the
GW, we randomly\footnote{Although this may seem somewhat arbitrary, it is no
more arbitrary than choosing a Gaussian distribution as in the last section. We
have performed the numerical integration below for many different choices of
$\bm\chi_0$, and the results are qualitatively similar.} choose
$\bm\chi^{(\omega_n)}=\bm\chi_0$. In order to compare  differing amplifications
for each parity, we use the same $\bm\chi_0$ for both parities. We then
integrate Eqs.~(\ref{MEGI1})~--~(\ref{alleqnsevenodd}) out to some
$r=r_{\mbox{max}}$ for each QNM frequency $\omega_n$. Then, for each variable
at $r=r_{\mbox{max}}$ we can simply add up the QNMs. This then gives a good
approximation to the time decay of the signal as it passes $r=r_{\mbox{max}}$
after $t\gapp t _{\mbox{max}}=r_{\mbox{max}}-r_0
+2m\ln\bras{\bra{r_{\mbox{max}}-2m}/\bra{r_0-2m}}$~\citep{andersson2,QNM}. We
use the first twelve QNM frequencies as tabulated in~\citet{NS} for
$\sigma_n=\tfrac{1}{2}\phi r\omega_n$~-- see Eq.~(\ref{omdef}).

\begin{figure}[ht!]
\begin{center}
\includegraphics[width=0.7\textwidth]{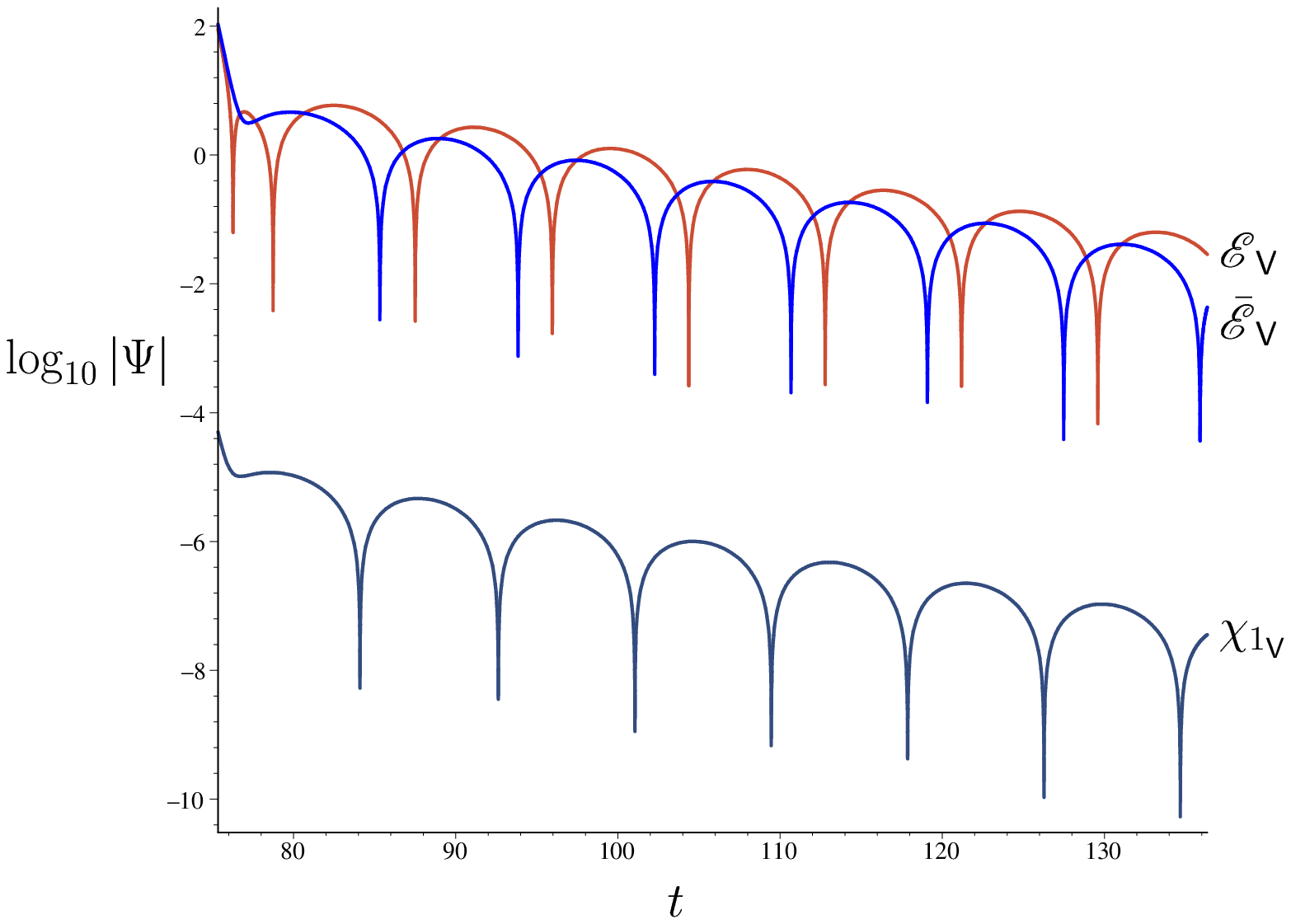}
\end{center}
\caption{Late time behavior showing the temporal evolution of the
generated EM waveforms for an observer at $r=65m$, using a quasi-normal mode
expansion. Choosing boundary conditions at $r_0=2.05m$, the QNM sum will
approximate the true solution for this observer after $t\simeq76m$. We see that
the EM waveform is substantially larger than the interaction variables, for
which we have shown the largest, $\chi_{1\V}$. The differing amplification of
the electric field for each parity is also not insubstantial, with the even
parity case being larger than the odd, in this case, demonstrating polarisation
of the induced EM radiation.
\label{QNMring}}
\end{figure}

In Figure~\ref{QNMring} we show a typical result of this integration, for an
observer situated at $r\simeq65$, with $r_0=2.05$. The generated electric field
is shown, for both parities, as is the largest interaction variable, which is
$\chi_{1\V}$ in this case. The units of the graph are arbitrary: dividing each
variable by $|\bm\chi_0|$ (to make each variable dimensionless), say, will
merely shift all the curves up or down. At large distances from the source, the
behaviour of the fields can be represented as an amplitude over a potential of
the distance function. In the case of the gravitational wave, the fall-off
scales like $1/r$, while for a spherical electromagnetic wave it behaves as
$1/r$. At the same time, the background magnetic dipole field has
$1/r^3$-dependence. We can therefore normalise with the respect to the fall-off
of the field strengths, in order to get a scale-invariant form of the
amplification. In the situation given in Fig.~\ref{QNMring}, normalising the
curve for $\chi_{1\V}$ raises it up by $3\log_{10}\bra{65/2.05}\sim4.5$. Hence,
at large distances from the source \emph{the scale-invariant amplification of
the EM radiation is over two orders of magnitude larger the the magnitude of
the GW times the magnetic field strength}. At this distance from the source the
interaction is no longer taking place to a significant degree, implying that
this level of amplification is a generic feature. Note also from
Fig.~\ref{QNMring} that the amplification of the electric field parities is
different, implying that the EM wave is polarised.

\subsection{Estimates}

To estimate the implications of this amplification we have found,
consider the case of a compact object such as a BH or neutron
star. The interaction between gravity waves and the magnetic field
is quantified by the variable $\chi_i \sim h_i B_i$, where $h_i$
is the amplitude of the gravitational wave at the onset of the
interaction, $B_i \sim B_\text{s}(r_\text{s}/r_i)^3$ is the field
strength at the distance $r_i$ from the compact object with
`radius' (e.g., surface of a neutron star or BH horizon)
$r_\text{s}$ and surface magnetic field $B_\text{s}$. The
interaction produces an EM signal $E_\text{out}$ which is
typically two orders of magnitude larger at $r_\text{cut-off}$
than the original perturbation $\chi_1$ at $r_\text{cut-off}$,
where the interaction effectively switches off. From Fig. 3 we
extrapolate that at $r_\text{cut-off} \sim 40 m$ we have roughly
$\chi_\text{cut-off}\sim 10^{-2.5}\chi_i$, leading to an induced
electric field strength
\be
E_\text{out}\sim  3\times 10^{8}
h_i\bra{\frac{B_\text{s}}{1\,\mathrm{T}}}\bra{\frac{r_\text{s}}{r_{i}}}^3
\mathrm{Vm^{-1}}. \label{Eout}
\ee
The induced signal attenuates inversely with distance $D$ outside the
interaction region, e.g. $r\gtrsim 40m$. At a distance $D$ from the source, the
spectral energy flux $\Phi_\omega = (\tfrac12\epsilon_0cE^2)/\omega$ can then
be calculated to be
\be
\Phi_\omega \sim 1.4\times 10^{6}
h_i^2\bra{\frac{B_\text{s}}{1\,\mathrm{T}}}^2
\bra{\frac{r_\text{s}}{r_i}}^6 \bra{\frac{r_\text{cut-off}}{100\,\mathrm{km}}}^2
\bra{\frac{10 \,\mathrm{kpc}}{D}}^2 \bra{\frac{1\, \mathrm{kHz}}{\omega}}
\mathrm{Jy}. \label{Jansky}
\ee
As an example, consider a magnetar of mass $m=1.5 \,M_{\odot}$ with radius
$r_\text{s} = 9$ km, e.g. twice its Schwarzschild radius, and take $r_i = 4
\,r_\text{s}$ and $r_\text{cut-off}=90$ km, assuming a magnetic field strength
$B_\text{s}$ in the range of $10^5$ to $10^{10}$ T. An occurring instability
such as a supernova explosion or a bar mode instability is likely to produce a
GW with $h_i\sim10^{-3}$ and frequency $\omega$ of about $1-10$
kHz~\citep{toprev}, which is also the frequency of the induced EM wave. This
leads to $E_\text{out}
\sim 5\times10^{8} - 10^{13}\, \mathrm{Vm}^{-1}$.  If such an event happened
within our galaxy ($D \sim 10$ kpc), $\Phi_\omega\sim 10^{5} -  10^{15}$ Jy, if
we assume that 10\% of the signal's energy undergoes mode conversion shifting
the frequency up to $30-300$ kHz~\citep{MBD}. To achieve a higher detection
rate, one has to gather events from a farther distance. Events within the Virgo
Cluster ($D
\sim 15$ Mpc) would have flux $\Phi_f\sim10^{-1} -
10^{9}$ Jy. The proposed radio telescope Astronomical Low Frequency Array
\citep{alfa} is expected to operate in the range from 30 kHz to 30 MHz, with
minimum detection level of 1000 Jy, making such events an exciting possibility
for indirect gravitational wave detection.


\section{Conclusions}\label{sec5}

We have investigated the scenario of GW  around a Schwarzschild BH interacting
with a strong, static, magnetic field. This interaction produces a stream of EM
radiation mirroring the BH ringdown, with a stronger amplitude than one may
expect from estimates of the interaction in flat space, due to non-linear
amplification in the vicinity of the photon sphere. This interaction may play
an important role in GRBs and perhaps some SN events, in addition to neutron
star physics, and may be a useful mechanism to aid in GW detection.

We converted the Einstein-Maxwell equations into a linear, gauge-invariant
system of differential equations by utilising the 1+1+2 covariant approach to
perturbations of Schwarzschild. We also introduced a set of second order
`interaction' variables to aid in simplifying the derivation, and a new
variable for the magnetic field, both of which made the system of equations
manifestly gauge-invariant.  It was then a simple matter to convert the system
of equations into wave equations for integration as an initial value problem,
or as a harmonically decomposed (in time) system of first-order ordinary
differential equations, which could then be integrated using a BH quasi-normal
mode expansion, an important approximation method for late time behaviour. We
integrated the system of equations using both of these techniques.

A key point of this paper was to set up a suitable formalism to study this
GW-B interaction around a BH, and to put the equations into a suitable
gauge-invariant form for numerical integration. The next step is to include a
plasma, as various plasma instabilities could be induced by such process,
making detection of this sort of induced radiation a genuine possibility.
This will also help model some of the relativistic effects which take place
after a SN explosion. In fact, EM waves in a plasma in an \emph{exact}
Schwarzschild spacetime are pretty complicated and unexpected~\citep{DT}, so
it is an interesting question in its own right to ask what happens when GW
are thrown into the mix.

\acknowledgments

This project was financially supported by NRF (South Africa) and Sida (Sweden).
CAC would like to thank the Department of Electromagnetics at Chalmers
University of Technology for their hospitality while part of this work was
undertaken. We would like to thank Richard Barrett and Marco Bruni  for useful
discussions.

\appendix
\section{Spherical harmonics}

Here we briefly review the spherical harmonic expansion, developed
in~\citet{CB} appropriate to the formalism, for easy reference. These allow us
to remove all $\delta$-derivatives from the equations. Note that all
functions and relations below are defined in the background only; we only
expand second-order variables, or first-order variables which form part of a
quadratic second-order variable so zeroth-order equations are sufficient.

We introduce spherical harmonic functions $Q=Q^{(\ell,m)}$, with
$m=-\ell,\cdots,\ell$, defined on the background, such that
\be
\delta^2 Q = -\ell\bra{\ell+1}r^{-2} Q,~~~\hat Q=0=\dot Q.\label{SH}
\ee
We also need to expand vectors and tensors in spherical harmonics. We
therefore define the \emph{even} (electric) parity vector spherical harmonics
for $\ell\geq1$ as
\begin{mathletters}
\ba
Q_a^{(\ell)}=r\delta_a Q^{(\ell)} ~~~\Rightarrow ~~~
\hat Q_a=0=\dot Q_a,~~\delta^2Q_a=\bra{1-\ell\bra{\ell+1}}r^{-2}Q_a;
\ea
where the $(\ell)$ superscript is implicit, and we define \emph{odd}
(magnetic) parity vector spherical harmonics as
\ba
\bar Q_a^{(\ell)}=r\lc_{ab}\delta^b Q^{(\ell)}~~~\Rightarrow~~~
 \hat{\bar{Q}}_a=0=\dot{\bar{Q}}_a,~~~\delta^2\bar
Q_a=\bra{1-\ell\bra{\ell+1}}r^{-2}\bar Q_a.
\ea
\end{mathletters}
Note that $\bar Q_a=\lc_{ab}Q^b\Leftrightarrow Q_a=-\lc_{ab}\bar Q^b$, so
that $\lc_{ab}$ is a parity operator. The crucial difference between these
two types of vector spherical harmonics is that $\bar Q_a$ is solenoidal, so
\be
\delta^a\bar Q_a=0,~~~\mbox{while}~~~
\delta^aQ_a=-\ell\bra{\ell+1}r^{-1} Q.
\ee
Note also that
\be
\lc_{ab}\delta^a  Q^b=0,~~~\mbox{and}~~~\lc_{ab}\delta^a\bar Q^b=\ell\bra{\ell+1}r^{-1} Q.
\ee
The harmonics are orthogonal: $Q^a\bar Q_a=0$ (for each $\ell$). Similarly we
define even and odd tensor spherical harmonics for $\ell\geq2$ as
\begin{mathletters}
\ba
Q_{ab} = r^2\delta_{\lb a}\delta_{b\rb}Q,~~~\Rightarrow~~~\hat
Q_{ab}=0=\dot Q_{ab},~~~
\delta^2Q_{ab} = \bras{\phi^2-3\E-\ell\bra{\ell+1}r^{-2}}Q_{ab};
\ea
and
\ba
\bar Q_{ab} = r^2\lc_{c\lb
a}\delta^c\delta_{b\rb}Q,~~~\Rightarrow ~~~\hat{\bar Q}_{ab}= 0
=\dot{\bar Q}_{ab},~~~
\delta^2\bar Q_{ab} =
\bras{\phi^2-3\E-\ell\bra{\ell+1}r^{-2}}\bar Q_{ab},
\ea
\end{mathletters}
which are orthogonal: $Q_{ab}\bar Q^{ab}=0$, and are parity inversions of one
another: $Q_{ab}=-\lc_{c\lb a}\bar Q_{b\rb}^{~~c}\Leftrightarrow
\bar Q_{ab}=\lc_{c\lb a} Q_{b\rb}^{~~c}$.

We can now expand any second-order scalar ${\Psi}$ in terms of these
functions as
\be
{\Psi}=\sum_{\ell=0}^{\infty}\sum_{m=-\ell}^{m=\ell} {\Psi}\sfS^{(\ell,m)}
Q^{(\ell,m)} = {\Psi}\sfS Q,
\ee
where the sum over $\ell$ and $m$ is implicit in the last equality. The
$\mathsf{S}$ subscript reminds us that ${\Psi}$ is a scalar, and that a
spherical harmonic expansion has been made. Due to the spherical symmetry of
the background, $m$ never appears in any equation so we can just ignore it.
Any second-order vector ${\Psi}_a$ can now be written
\be
{\Psi}_a=\sum_{\ell=1}^{\infty} {\Psi}^{(\ell)}\V Q_a^{(\ell)}+\bar
{\Psi}^{(\ell)}\V\bar Q_a^{(\ell)}={\Psi}\V Q_a+\bar {\Psi}\V\bar Q_a.
\ee
Again, we implicitly assume a sum over $\ell$ in the last equality, and the
$\mathsf{V}$ reminds us that ${\Psi}^a$ is a vector expanded in spherical
harmonics.
 Any second-order tensor may be also be expanded
\be
{\Psi}_{ab}=\sum_{\ell=2}^{\infty} {\Psi}\T^{(\ell)}Q_{ab}^{(\ell)}+\bar
{\Psi}\T^{(\ell)}\bar Q_{ab}^{(\ell)}={\Psi}\T Q_{ab}+\bar {\Psi}\T\bar
Q_{ab}.
\ee
Further useful identities are to be found in~\citet{CB}.

\end{document}